\documentclass[journal]{IEEEtran}
\usepackage{float}
\usepackage{amsmath,amsfonts,amssymb}
\usepackage{graphicx}
\usepackage{cite}
\usepackage{multirow}
\usepackage{algorithm}
\usepackage{algorithmic}
\usepackage{subfigure}
\usepackage{multirow}
\usepackage{url}
\usepackage{color}
\usepackage{array}
\usepackage{caption}

\graphicspath{{./figsJournal/}}

\IEEEoverridecommandlockouts

\makeatletter
\newcommand*{\rom}[1]{\expandafter\@slowromancap\romannumeral #1@}
\makeatother

\newcommand{\upa}{\text{U}}

\renewcommand{\ast}{\circledast}

\newtheorem{lemma}{Lemma}

\newtheorem{remark}{Remark}

\newcommand{\Qrf}{Q_{\text{RF}}}

\newcommand{\Mrf}{M_{\text{RF}}}
\newcommand{\tD}{\tilde{D}}
\newcommand{\ts}{\tilde{s}}
\newcommand{\ty}{\tilde{y}}
\newcommand{\tphi}{\tilde{\phi}}
\newcommand{\ttheta}{\tilde{\theta}}

\newcommand{\sinc}{\mathrm{sinc}}

\newcommand{\xE}{\mathbb{E}}

\newcommand{\diag}{\mathrm{diag}}

\begin{document}
\title{Millimeter Wave MIMO with Lens Antenna Array: A New Path Division Multiplexing Paradigm}
\author{Yong~Zeng and Rui~Zhang
\thanks{Y. Zeng is with the Department of Electrical and Computer Engineering, National University of Singapore (e-mail: elezeng@nus.edu.sg).}
\thanks{R. Zhang is with the Department of Electrical and Computer Engineering, National University of Singapore (e-mail: elezhang@nus.edu.sg). He is also with the Institute for Infocomm Research, A*STAR, Singapore.}
}

\maketitle

\begin{abstract}
Millimeter wave (mmWave) communication over the largely unused  mmWave spectrum is a promising technology for the fifth-generation (5G) cellular systems. To compensate for the severe path loss in mmWave communications, large antenna arrays are generally used at both the transmitter and receiver to achieve significant  beamforming gains. However, the high hardware and power consumption cost due to the large number of radio frequency (RF) chains required renders  the traditional beamforming method impractical for mmWave systems. It is thus practically valuable to achieve the large-antenna gains, but with only limited number of RF chains for mmWave communications. To this end, we study in this paper a new  \emph{lens antenna array} enabled  mmWave multiple-input multiple-output (MIMO) communication system. We first show that the array response of the proposed lens antenna array at the receiver/transmitter follows a ``sinc'' function, where the antenna with the peak response is determined by the angle of arrival (AoA)/departure (AoD) of the received/transmitted signal.  By exploiting this unique property of lens antenna arrays along with the multi-path sparsity of mmWave channels, we propose a novel low-cost and capacity-achieving MIMO transmission scheme, termed \emph{orthogonal path division multiplexing (OPDM)}.  With OPDM, multiple data streams are simultaneously transmitted in parallel  over different propagation paths with simple \emph{per-path} signal processing at both the transmitter and receiver. For channels with insufficiently separated AoAs and/or AoDs, we also propose a simple \emph{path grouping} technique with group-based small-scale MIMO processing to mitigate the inter-path interference. Numerical results are provided to compare the performance of the proposed lens antenna arrays for mmWave MIMO system against that of conventional arrays, under different practical setups. It is shown that the proposed system  achieves significant throughput gain as well as complexity and hardware cost reduction, both making it an appealing new paradigm for mmWave MIMO communications.
\end{abstract}


\section{Introduction}
The fifth-generation (5G) wireless communication systems on the roadmap are expected to provide at least 1000 times capacity increase over the current 4G systems \cite{550}. To achieve this goal, various  technologies have been proposed and extensively investigated  during the past few years \cite{568}. Among others, wireless communication over the largely unused millimeter wave (mmWave) spectrum (say, 30-300GHz) is regarded as a key enabling technology for 5G and has drawn  significant interests recently (see \cite{566,567,486,569} and the references therein). Existing mmWave communication  systems are  designed mainly for short-range line-of-sight (LOS) indoor applications, e.g., wireless personal area networking (WPAN) \cite{571} and wireless local area networking (WLAN) \cite{581}. While recent measurement results  have shown that, even in non-line-of-sight (NLOS) outdoor environment, mmWave signals with  satisfactory strengths can be received up to 200 meters \cite{565,594}, which indicates that mmWave communications may also be feasible for future cellular networks with relatively small cell coverage.

MmWave signals generally experience orders-of-magnitude more path loss than those at much lower frequency in existing cellular systems. On the other hand, their smaller wavelengths make it practically feasible to pack a large number of antennas with reasonable form factors at both the transmitter and receiver, whereby efficient MIMO (multiple-input multiple-output) beamforming techniques can be applied to achieve highly directional communication to compensate for the severe path loss \cite{594,580,582,591}. However, traditional MIMO beamforming is usually implemented digitally at baseband and thus requires one dedicated  radio frequency (RF) chain for each transmit/receive antenna, which may not be feasible in mmWave systems due to the high hardware  and power consumption cost of the large number of RF chains required. To reduce the cost and yet achieve the high array gain, {\it analog beamforming} has been proposed for mmWave communications \cite{574,575,573}, which can be implemented via phase shifters in the RF frontend, and thus requires only one RF chain  for the entire transmitter/receiver. Despite of the notable cost reduction, analog beamforming usually incurs significant performance loss  due to the constant-amplitude beamformer constraint imposed by the phase shifters, as well as its inability to perform spatial multiplexing for high-rate transmission. To enable spatial multiplexing,  {\it hybrid analog/digital precoding} has been recently proposed \cite{576,577,578,593,579,592}, where the precoding is implemented in two stages with a  baseband digital precoding using a limited number of RF chains followed by a RF-band analog processing through a network of phase shifters.  Since the hybrid precoding  in general requires a large number of phase shifters,  antenna subset selection has been proposed in \cite{583} by replacing the phase shifters with switches. However, antenna selection may cause significant performance degradation due to the limited array gains resulted \cite{369,370}, especially in highly correlated MIMO channels as in mmWave systems.

Besides, another promising line of research for mmWave or large MIMO systems aims to reduce signal processing complexity and RF chain cost without notable performance degradation by utilizing advanced antenna designs, such as the \emph{lens antenna array} \cite{584,553,456,485,585}. As shown in Fig.~\ref{F:lensArray}, a lens antenna array is in general composed of two main components: an electromagnetic (EM) lens and a matching antenna array with elements located in the focal region of the lens. Generally speaking, EM lenses can be implemented via three main technologies: i) the dielectric lenses made of dielectric materials with carefully designed front and/or rear surfaces \cite{488,376}; ii) the traditional planar lenses consisting of arrays of transmitting and receiving antennas connected via transmission lines with variable lengths \cite{560,561}; and iii) the modern planar lenses composed of sub-wavelength periodic inductive and capacitive structures \cite{554,559}. Regardless of the actual implementation methods, the fundamental principle of EM lenses is to provide variable phase shifting for EM rays at different points on the lens aperture so as to achieve \emph{angle of arrival (AoA)/departure (AoD)-dependent energy focusing}, i.e., a receiving (transmitting) lens antenna array is able to focus (steer) the incident (departure) signals with sufficiently separated AoAs (AoDs) to (from) different antenna subsets.   In \cite{553}, the concept of beamspace MIMO communication is introduced, where the lens antenna arrays are used to approximately transform the signals in  antenna space to beamspace, which has much lower dimensions, to significantly reduce the number of RF chains required. However, the studies in \cite{553} focus on the LOS mmWave channels, where spatial multiplexing is possible only for very short transmission range  (e.g.  a few meters) and/or extremely large antenna apertures.  In a parallel work \cite{485},  the lens antenna array is applied to the massive MIMO cellular system with large number of antennas at the base station (BS) \cite{373,374,497}, which is shown to achieve significant performance gains as well as cost reduction as compared to the conventional arrays without lens. However, the result in \cite{485} is only applicable for the single-input multiple-output (SIMO) uplink transmission, instead of the more general setup with lens antenna arrays applied at both the transmitter and receiver. Moreover, neither \cite{553} nor \cite{485} fully explores the characteristics of mmWave channels, such as the multi-path sparsity \cite{486} due to limited scattering  and the frequency selectivity in broadband transmission.

In this paper, we study the mmWave  MIMO communication where both the transmitter and receiver are equipped with lens antenna arrays. Due to the AoA/AoD-dependent energy focusing,  in mmWave systems with limited number of multi-paths, the signal power is generally focused on only a small subset of the antenna elements in the lens array; as a result, antenna selection can be applied to significantly reduce the RF chain cost, yet without notably comprising the system performance, which is in sharp contrast to the case of applying antenna selection with the conventional arrays \cite{369,370}. Furthermore, for mmWave channels with sufficiently separated AoAs/AoDs, different signal paths  can be differentiated in the spatial domain with the use of the lens antenna array. Therefore, the detrimental multi-path  effect in wide-band communications, i.e., the inter-symbol interference (ISI), can be easily alleviated in the lens array MIMO systems, without the need of sophisticated ISI mitigation techniques such as equalization, spread spectrum, or multi-carrier transmission \cite{56}.
In fact, in the favorable scenario where the AoAs/AoDs are sufficiently separated, the lens MIMO system can be shown to be equivalent to a set of parallel additive white Gaussian noise (AWGN) sub-channels, each corresponding to one of the multi-paths, for both narrow-band and wide-band communications. Thus, multiple data streams can be simultaneously  multiplexed and transmitted over these sub-channels in parallel, each over one of the multi-paths with simple per-path processing.  We term this new MIMO spatial multiplexing scheme enabled by the lens antenna array as {\it orthogonal path division multiplexing} (OPDM), in contrast to the conventional multiplexing techniques over orthogonal time or frequency.\footnote{Note that OPDM also differs from the conventional sectorized antenna and space-division-multiple-access (SDMA) techniques. Although they similarly exploit the different AoAs/AoDs of multiuser/multi-path signals, the former achieves spatial signal separation only in a coarse scale (say, 120 degrees with a 3-sector antenna array), while the latter obtains finer spatial resolution but with sophisticated beamforming/precoding. In contrast, with the proposed OPDM, high spatial resolution is  achieved without the need of complex array signal processing.}         We summarize the main contributions of this paper as follows.
\begin{itemize}
\item First, we present the  array configuration for the  proposed lens antenna array in detail, and derive its corresponding array response. Our result shows that, different from the conventional arrays whose response is generally given by phase shifting across the antenna elements, the array response for the lens antennas follows  a ``sinc'' function, where the antenna with peak response is determined by the AoA/AoD of the received/transmitted signal. This analytical result is consistent with that reported in prior works based on simulations \cite{376,585} or experiments \cite{554}. With the derived array response, the channel model for the   lens MIMO system is obtained, which is compared with that of a benchmark system using the conventional uniform planar arrays (UPAs).
\item Next, to obtain fundamental limit and draw insight,  we consider the so-called ``ideal'' AoA/AoD environment, where the signal power of each multi-path is focused on one single element of the lens array at the receiver/transmitter. We show that the channel capacity in this case is achieved by the novel   OPDM scheme, which can be easily implemented by antenna selection with only $L$ transmitting/receiving RF chains, with $L$ denoting the number of multi-paths. Notice that $L$ is usually much smaller than the number of transmitting/receiving antennas in mmWave MIMO channels due to the multi-path sparsity.  We further compare the lens array based mmWave MIMO system with that based on the conventional UPAs, in terms of capacity performance as well as signal processing complexity and RF chain cost.
\item Finally, the mmWave lens MIMO is studied under the practical setup with multi-paths of arbitrary AoAs/AoDs.
     We propose a low-complexity transceiver design based on path-division multiplexing (PDM),  applicable for both narrow-band and wide-band communications,  with {\it per-path} maximal ratio transmission (MRT) at the transmitter and maximal ratio combining (MRC)/minimum mean square error (MMSE) beamforming at the receiver. We analytically show that in the case of wide-band communications, the proposed design achieves perfect ISI rejection if {\it either} the AoAs or AoDs (not necessarily both) of the multi-path signals are sufficiently separated, which usually holds in practice. Moreover, for cases with insufficiently separated AoAs and/or AoDs, we propose a simple {\it path grouping} technique with group-based small-scale MIMO processing to mitigate the inter-path interference.
\end{itemize}

It is worth pointing out that there has been an upsurge of interest recently in exploiting the angular domain of multi-path/multiuser signals in the design of massive MIMO systems. For example, by utilizing the fact that there is  limited angular spread for signals sent from the mobile users, the authors in \cite{587} propose a channel covariance-based pilot assignment strategy to mitigate the pilot contamination problem in multi-cell massive MIMO systems. Similarly in \cite{588,589}, an AoA-based user grouping technique is proposed, which leads to the so-called joint spatial division and multiplexing scheme that makes   massive MIMO  also possible for frequency division duplexing (FDD) systems due to the significantly reduced channel estimation overhead after user grouping. In \cite{590}, an OFDM (orthogonal frequency division multiplexing)  based beam division multiple access scheme is proposed for massive MIMO systems by simultaneously serving users with different beams at each frequency sub-channel. In this paper, we also exploit the different AoAs/AoDs of multi-path signals for complexity and cost reduction in mmWave MIMO systems, by utilizing the novel lens antenna arrays at both the transmitter and receiver. %, for both narrow-band and wide-band communications.

 The rest of this paper is organized as follows. Section~\ref{sec:channelModel} presents the array architecture as well as the array response function of the proposed lens antenna, based on which the MIMO channel model for mmWave communications is derived. The benchmark system using the conventional UPAs is also presented. In Section~\ref{sec:idealAoAAoD}, we consider the case of ``ideal'' AoA/AoD environment to introduce OPDM and demonstrate the great advantages of applying lens antenna arrays over conventional UPAs in mmWave communications. In Section~\ref{sec:arbAoAAoD}, the practical scenario with arbitrary AoAs/AoDs is considered, where a simple transceiver design termed PDM applicable for both narrow-band and wide-band communications is presented, and  a path grouping technique is proposed to further improve the performance. Finally, we conclude the paper and point out   future research directions in Section~\ref{sec:Conclu}.

\emph{Notations:} In this paper, scalars are denoted by italic letters. Boldface lower- and upper-case letters denote vectors and matrices, respectively. $\mathbb{C}^{M\times N}$ denotes the space of $M\times N$ complex-valued matrices, and $\mathbf{I} $ represents an    identity matrix.
For an arbitrary-size matrix $\mathbf{A}$,  its complex conjugate, transpose, and Hermitian transpose are denoted by $\mathbf A^*$, $\mathbf{A}^{T}$, and $\mathbf{A}^{H}$, respectively. %For a square Hermitian matrix $\mathbf{S}$, $\mathrm{Tr}(\mathbf{S})$ denotes its trace, while $\lambda_{\max}(\mathbf S)$ and $\mathbf v_{\max}(\mathbf S)$ denote its largest eigenvalue and the corresponding eigenvector, respectively.
  For a vector $\mathbf a$, $\|\mathbf a\|$ denotes its Euclidean norm, and $\diag{(\mathbf a)}$ represents a diagonal matrix with the diagonal elements given in $\mathbf a$. For a non-singular square matrix $\mathbf S$, its matrix inverse is denoted as $\mathbf S^{-1}$. The symbol $j$ represents the imaginary unit of complex numbers, with $j^2=-1$. %$[\mathbf A]_{nm}$ denotes the $(n,m)$-th element of matrix $\mathbf A$, and $[\mathbf v]_{1:K}$ denotes a vector consisting of the first $K$ elements of vector $\mathbf v$.
 The notation $\ast$ denotes the linear convolution operation. $\delta(\cdot)$ denotes the Dirac delta function, and $\sinc(\cdot)$ is the ``sinc'' function defined as $\sinc(x)\triangleq \sin(\pi x)/(\pi x)$. For a real number $a$, $\lfloor a \rfloor$ denotes the largest integer no greater than $a$, and $\mathrm{round}(a)$ represents the nearest integer of $a$. Furthermore, $U[a, b]$ represents the uniform distribution in the interval $[a, b]$. $\mathcal{N}(\boldsymbol \mu, \mathbf C)$ and $\mathcal{CN}(\boldsymbol \mu, \mathbf C)$   denote the real-valued Gaussian and the circularly symmetric complex-valued Gaussian (CSCG) distributions with mean $\boldsymbol \mu$ and covariance matrix $\mathbf C$, respectively. For a set $\mathcal S$, $|\mathcal S|$ denotes its cardinality. Furthermore, $\mathcal S_1 \cap \mathcal S_2$ and $\mathcal S_1 \cup \mathcal S_2$ denote the intersection and union of sets $S_1$ and $S_2$, respectively.

 \section{System Description and Channel Model}\label{sec:channelModel}
 \subsection{Lens Antenna Array}\label{sec:lensArray}
\begin{figure}
\centering
\includegraphics[scale=0.6]{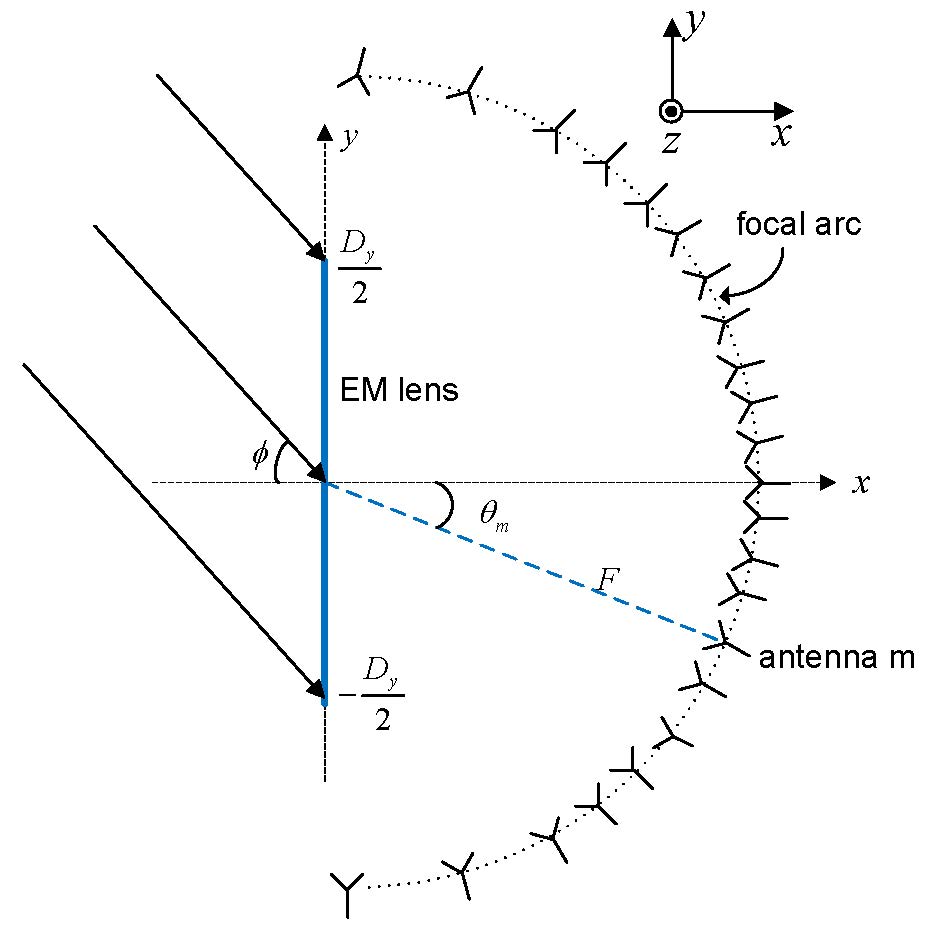}
\caption{The schematic diagram of a lens antenna array with an incident uniform plane wave with AoA $\phi$.}\label{F:lensArray}
\end{figure}

  A lens antenna array in general consists of an EM lens and an antenna array with elements located in the focal region of the lens.
   Without loss of generality, we assume that a planar EM lens with negligible thickness and of size $D_y\times D_z$ is placed on the y-z plane and centered at the origin, as shown in Fig.~\ref{F:lensArray}. By considering only the azimuth AoAs and AoDs,\footnote{For simplicity, we assume that the elevation AoAs/AoDs are all zeros, which is practically valid if the height difference between the transmitter and the receiver is much smaller than their separation distance.} the array elements are assumed to be placed on the {\it focal arc} of the lens, which is defined as a semi-circle  around the lens's center in the azimuth plane (i.e., x-y plane shown in Fig.~\ref{F:lensArray}) with radius $F$, where $F$ is known as the focal length of the lens. Therefore, the antenna locations  relative to the lens center can be parameterized as $B_m(x_m=F\cos \theta_m, y_m=-F\sin \theta_m, z_m=0)$, where $\theta_m\in [-\pi/2, \pi/2]$ is the angle of the $m$th antenna element relative to the x-axis, $m\in \mathcal{M}$, with $\mathcal{M}\triangleq \{0, \pm 1, \cdots, \pm(M-1)/2\}$ denoting the set of antenna indices and $M$ representing the total number of antennas. Note that we have assumed that $M$ is an odd number for convenience. Furthermore, we assume the so-called critical antenna spacing, i.e., the antenna elements are deployed on the focal arc so that $\{\tilde \theta_m \triangleq \sin \theta_m\}$ are equally spaced in the interval $[-1, 1]$ as
  \begin{align}
\tilde \theta_m = \frac{m}{\tD}, \ m\in \mathcal{M}, \label{eq:sinThetam}
  \end{align}
  where $\tD\triangleq D_y/\lambda$ is the effective lens dimension along the azimuth plane, with $\lambda$ denoting the carrier wavelength. It follows from \eqref{eq:sinThetam} that $M$ and $\tD$ are related via $M=1+\lfloor 2\tD \rfloor$, i.e., more antennas should be deployed for larger lens dimension $\tD$. It is worth mentioning that with the array configuration specified in \eqref{eq:sinThetam}, antennas are more densely deployed in the center of the array than those on each of the two edges.

 We first study the receive array response by assuming that the lens antenna array is illuminated by a uniform plane wave with AoA $\phi$, as shown in Fig.~\ref{F:lensArray}. Denote by $x_0(\phi)$ the impinging signal at the reference point (say, the lens center) on the lens aperture, and $r_m(\phi)$ the resulting signal received by the $m$th element of the antenna array, $m\in \mathcal M$. The array response vector $\mathbf a(\phi)\in \mathbb{C}^{M\times 1}$, whose elements are defined by the ratio $a_m(\phi)\triangleq r_m(\phi)/x_0(\phi)$, can then be obtained in the following lemma.

\begin{lemma}\label{lemma:response}
For the lens antenna array with critical antenna spacing as specified in \eqref{eq:sinThetam}, the receive array response vector $\mathbf a(\phi)$ as a function of the AoA $\phi$ is given by
\begin{align}
a_m(\phi) = \sqrt{A} \sinc(m-\tD \tphi),\ m\in \mathcal{M}, \label{eq:am}
\end{align}
where $A\triangleq D_yD_z/\lambda^2$ is the effective aperture of the EM lens, and $\tphi\triangleq \sin \phi\in [-1, 1]$ is referred to as the {\it spatial frequency} corresponding to the AoA $\phi$.
\end{lemma}
\begin{IEEEproof}
Please refer to Appendix~\ref{A:lensArray}.
\end{IEEEproof}

\begin{figure}
\centering
\includegraphics[scale=0.6]{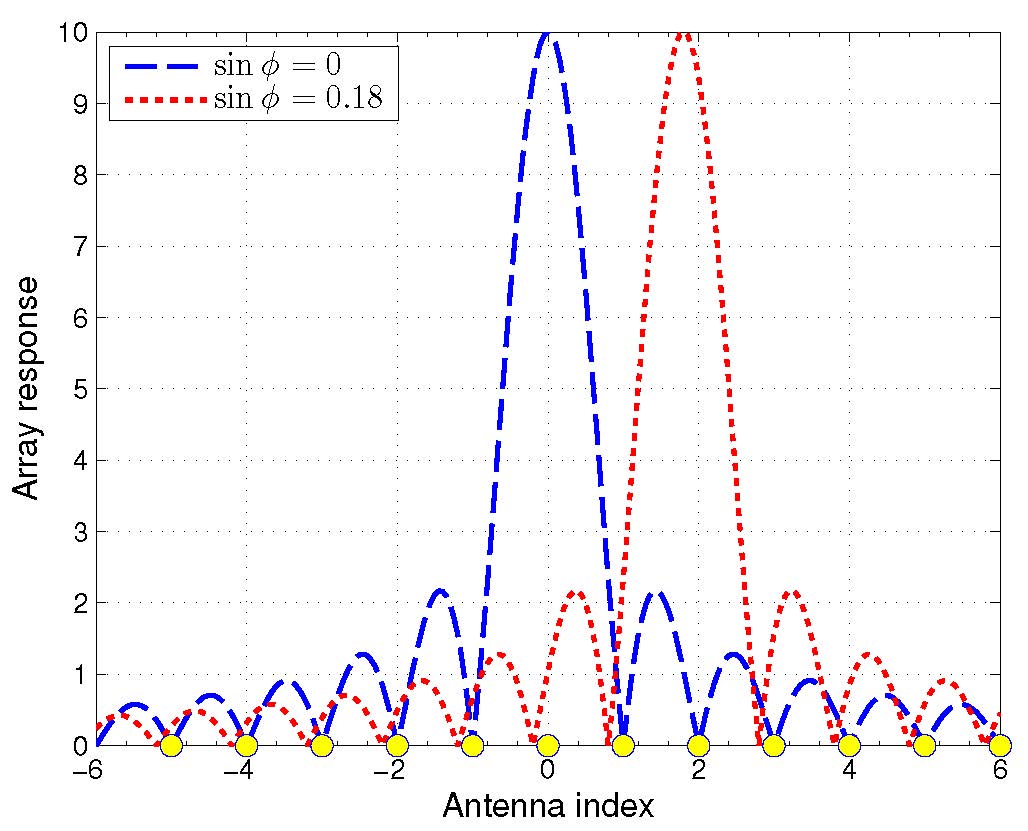}
\caption{Array response of a lens antenna array with $A=100$ and $\tD=10$ for two different AoAs.}\label{F:ArrayResponse}
\end{figure}

Different from the traditional antenna arrays without lens, whose array responses are generally given by the simple phase shifting across different antenna elements (see e.g. \eqref{eq:aRUPA} for the case of UPAs), the ``$\sinc$''-function array response in \eqref{eq:am} demonstrates the AoA-dependent energy-focusing capability of the lens antenna arrays, which is illustrated in Fig.~\ref{F:ArrayResponse}. Specifically, for any incident signal with a given AoA $\phi$, the received power is magnified by approximately $A$ times for the receiving antenna located in the close vicinity of the focal point $\tD\tphi$; whereas it is almost negligible for those antennas located far away from the focal point, i.e., antennas with $|m-\tD\tphi|\gg  1$. As a result, any two simultaneously received signals with sufficiently different AoAs $\phi$ and $\phi'$ such that $|\tphi -\tphi'|\geq 1/\tD$ can be effectively separated in the spatial domain, as illustrated in Fig.~\ref{F:ArrayResponse} assuming a lens antenna array with $A=100$ and $\tD=10$ for two AoAs with $\sin \phi=0$ and $0.18$, respectively. Thus, we term the quantity $1/\tD$  as the array's {\it spatial frequency resolution}, or approximately the AoA resolution for large $\tD$ \cite{553}.

On the other hand, since the EM lens  is  a passive device, reciprocity holds between the incoming and outgoing signals through it.
As a result, the transmit  response vector for steering a signal towards the AoD $\phi$ can be similarly obtained by Lemma~\ref{lemma:response}.

\subsection{Channel Model for MmWave Lens MIMO}\label{sec:modelLensMIMO}
\begin{figure}
\centering
\includegraphics[scale=0.48]{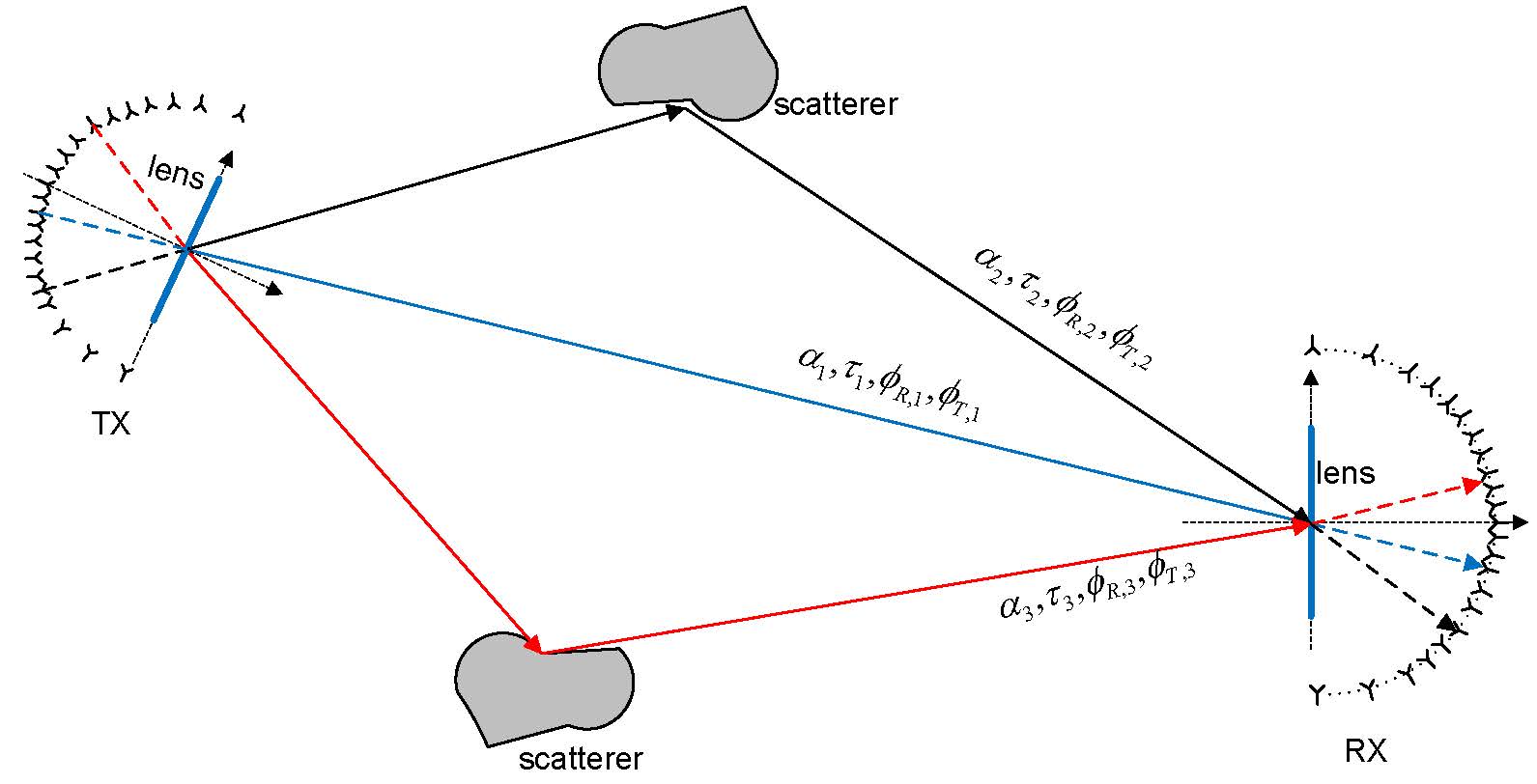}
\caption{A mmWave lens MIMO system in multi-path environment.}\label{F:lensMIMOMP}
\end{figure}

In this subsection, we present the channel model for the mmWave lens MIMO system, where both the transmitter and receiver are equipped with lens antenna arrays with $Q$ and $M$ elements, respectively, as shown in Fig.~\ref{F:lensMIMOMP}. Under the general multi-path environment, the channel impulse response can be modeled as
\begin{align}
\mathbf H(t) &= \sum_{l=1}^L \alpha_l  \mathbf a_{R}(\phi_{R,l}) \mathbf a_{T}^H(\phi_{T,l}) \delta(t-\tau_l),\label{eq:CIRMatrix}
\end{align}
where $\mathbf H(t)$ is an $M\times Q$ matrix with elements $h_{mq}(t)$ denoting the channel impulse response from transmitting antenna $q\in \mathcal{Q}$ to receiving antenna $m\in \mathcal M$, with $\mathcal{Q}$ and $\mathcal{M}$ respectively denoting the sets of the transmitting and receiving antenna indices as similarly defined in Section~\ref{sec:lensArray};  $L$ denotes the total number of significant multi-paths, which is usually small due to the multi-path sparsity  in mmWave channels \cite{486};  $\alpha_l$ and $\tau_l$ denote the complex-valued path gain and the delay for the $l$th path, respectively; %\footnote{We assume that the distances between the scatterers and the transmitter/receiver are much larger than the array dimensions, so that the signals can be well approximated as uniform plane waves and the propagation delays between the EM lens and the antenna elements are negligible as compared to $\{\tau_l\}$.}
  $\phi_{R,l}$ and $\phi_{T,l}$ are  the azimuth AoA and AoD for path $l$, respectively;  and $\mathbf a_{R}\in \mathbb{C}^{M\times 1}$ and  $\mathbf a_{T}\in \mathbb{C}^{Q\times 1}$ represent the array response vectors for the lens antenna arrays at the receiver and the transmitter, respectively. Note that in \eqref{eq:CIRMatrix}, we have assumed that the distances between the scatterers and the transmitter/receiver are much larger than the array dimensions, so that each multi-path signal can be well approximated as a uniform plane wave. % and the propagation delays between the EM lens and the antenna elements are negligible as compared to $\{\tau_l\}$

   Denote by $A_T$ and $A_R$ the effective lens apertures, and $\tD_T$ and $\tD_R$ the lens's effective azimuth dimensions at the transmitter and at the receiver, respectively. Based on Lemma~\ref{lemma:response}, the elements in the receive  and transmit  array response vectors $\mathbf a_{R}$ and $\mathbf a_{T}$ can be respectively expressed as
\begin{align}
a_{R,m}(\phi_{R,l})&=\sqrt{A_R} \sinc(m-\tD_R \tphi_{R,l}), \ m\in \mathcal{M}, \label{eq:arrayResponseRX}\\
a_{T,q}(\phi_{T,l})&=\sqrt{A_T} \sinc(q-\tD_T\tphi_{T,l}), \ q\in \mathcal{Q}, \label{eq:arrayResponseTX}
\end{align}
where $\tphi_{R,l}\triangleq \sin(\phi_{R,l})$ and $\tphi_{T,l}\triangleq \sin(\phi_{T,l})$ are the AoA/AoD spatial frequencies of the $l$th path. Without loss of generality, $\tphi_{R,l}$, $\tphi_{T,l}\in [-1,1]$ of the $L$ multi-paths can be expressed in terms of the spatial frequency resolutions associated with the receiving/transmitting arrays as
\begin{align}
\tphi_{R,l}=\frac{m_l+\epsilon_{R,l}}{\tD_R}, \ \tphi_{T,l}=\frac{q_l+\epsilon_{T,l}}{\tD_T}, \ l=1,\cdots, L, \label{eq:tphiMIMO}
\end{align}
where $m_l\in \mathcal{M}$ and $q_l\in \mathcal{Q}$ are integers given by  $m_l=\mathrm{round}(\tphi_{R,l}\tD_R)$ and $q_l=\mathrm{round}(\tphi_{T,l}\tD_T)$; and $\epsilon_{R,l}$ and $\epsilon_{T,l}$ are fractional numbers in the interval $[-1/2, 1/2]$. %For example, for a receiving lens antenna array with $\tD_R=10$ in a double-path environment with $\tphi_{R,1}=0.1$ and $\tphi_{R,2}=-0.18$, we have $m_1=1$, $m_2=-2$, $\epsilon_{R,1}=0$, and $\epsilon_{R,2}=0.2$.
Intuitively, $m_l$ (or $q_l$) in \eqref{eq:tphiMIMO} gives the receiving (transmitting) antenna index that is nearest to the focusing point corresponding to the AoA (AoD) of the $l$th path; whereas $\epsilon_{R,l}$ and $\epsilon_{T,l}$ represent the misalignment from the exact focusing point of the $l$th path signal relative to its nearest receiving/transmitting antenna.  By substituting \eqref{eq:tphiMIMO} into \eqref{eq:arrayResponseRX} and \eqref{eq:arrayResponseTX}, the channel impulse response in \eqref{eq:CIRMatrix} can be equivalently expressed as
\begin{equation}\label{eq:CIR2}
\begin{aligned}
h_{mq}(t)& = \sum_{l=1}^L \alpha_l \sqrt{A_RA_T}  \sinc (m-m_l-\epsilon_{R,l})\\
& \times \sinc(q-q_l-\epsilon_{T,l})\delta(t-\tau_l), \ m\in \mathcal{M}, q\in \mathcal{Q}.
\end{aligned}
\end{equation}
 Loosely speaking, \eqref{eq:CIR2} implies that the signal sent by the transmitting antenna with index $q=q_l$  will be  directed towards the receiver mainly along the $l$th path, and be mainly focused on the receiving antenna with index $m=m_l$, as illustrated in Fig.~\ref{F:lensMIMOMP}.

With the channel impulse response matrix $\mathbf H(t)$ given in \eqref{eq:CIRMatrix}, the baseband equivalent  signal received by the receiving lens antenna array can be expressed as
\begin{align}
\mathbf r(t)& = \mathbf H(t)\ast \mathbf x(t) + \mathbf z(t)\notag \\
&= \sum_{l=1}^L \alpha_l  \mathbf a_{R}(\phi_{R,l}) \mathbf a_{T}^H(\phi_{T,l})\mathbf x(t-\tau_l) + \mathbf z(t), \label{eq:rtMIMO}
\end{align}
where $\mathbf x(t)\in \mathbb{C}^{Q\times 1}$ denotes the signal sent from the $Q$ transmitting antennas, and $\mathbf z(t)\in \mathbb{C}^{M\times 1}$ represents the AWGN vector at the receiving antenna array. In the special case of narrow-band communications where the maximum excessive delay of the multi-path signals is much smaller than the symbol duration $T_s$, i.e., $\underset{l\neq l'}{\max}|\tau_l-\tau_{l'}|\ll T_s\approx 1/W$ with $W$ denoting the signal bandwidth, we have $\tau_l\approx \tau$ and $\mathbf x(t-\tau_l)\approx \mathbf x(t-\tau)$, $\forall l$. As a result, by assuming perfect time synchronization at the receiver, the general signal model for the wide-band communications in \eqref{eq:rtMIMO} reduces to
\begin{align}
\mathbf r(t) = \mathbf H \mathbf x(t) + \mathbf z(t),\label{eq:rtMIMONB}
\end{align}
where $\mathbf H=\sum_{l=1}^L \alpha_l  \mathbf a_{R}(\phi_{R,l}) \mathbf a_{T}^H(\phi_{T,l})$ denotes the narrow-band MIMO channel.

\begin{figure}%
\centering
\subfigure[Lens antenna array]{
\includegraphics[scale=0.6]{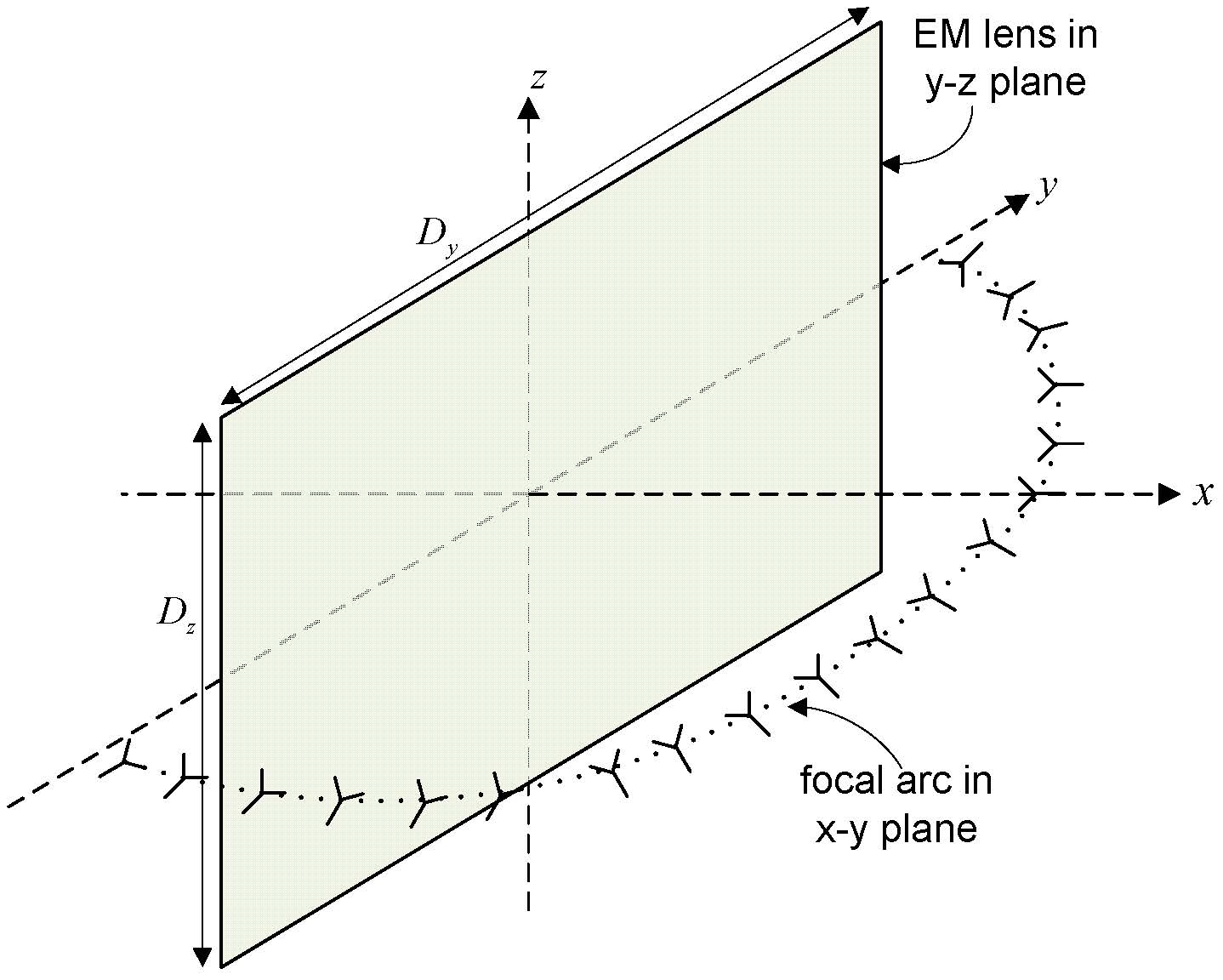}
\label{fig:subfigure1}
}
\quad
\vspace{-2ex}
\subfigure[Uniform planar array]{
\includegraphics[scale=0.6]{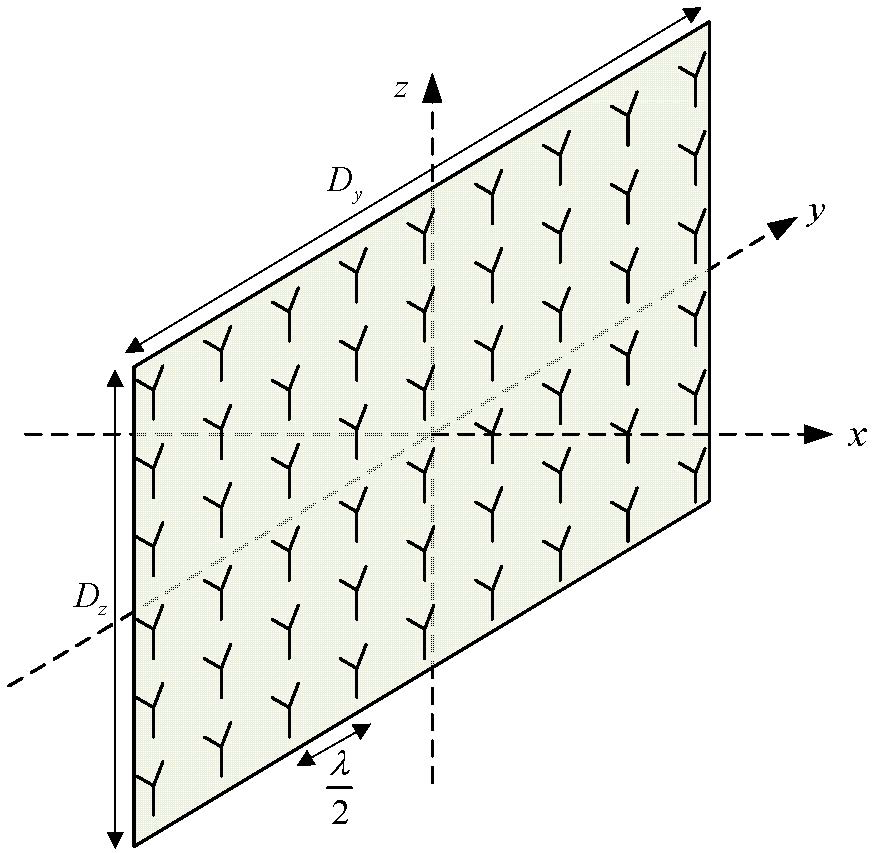}
%\end{minipage}%
\label{fig:subfigure2}
}
\caption{3D schematic diagrams of a lens antenna array versus an UPA with the same physical dimensions.}%
\label{F:antennas3D}
\end{figure}

%\begin{figure}%
%\centering
%\includegraphics[scale=0.5]{lensArray3D}
%\\
%\vspace{-5ex}
%(a)
%\\
%\includegraphics[scale=0.5]{UPA3D}
%%\end{minipage}%
%\\
%\vspace{-5ex}
%(b)
%\caption{3D schematic diagram for (a) lens antenna array; and (b) uniform planar array.}%
%\label{F:antennas3D}
%\end{figure}

\subsection{Benchmark System: MmWave MIMO with Uniform Planar Array}\label{sec:benchmark}
As a benchmark system for comparison, we   consider the mmWave communications in the traditional MIMO setup employing conventional antenna arrays without the EM lens. In particular, we assume that the transmitter and the receiver are both equipped with the UPAs with $Q_{\upa}$ and $M_{\upa}$ elements, respectively, with adjacent elements separated by distance $d_{\upa}=0.5\lambda$. For fair comparison, we assume that $Q_{\upa}$ and $M_{\upa}$ are designed such that the UPA has the same physical dimensions (or equivalently the same effective apertures $A_T$ and $A_R$) as the lens antenna of interest, as illustrated in Fig.~\ref{F:antennas3D}. Accordingly, it can be obtained that $Q_{\upa}=D_yD_z/d_{\upa}^2=4A_T>Q$ and $M_{\upa}=4A_R>M$, i.e., in general more antennas need to be deployed in the conventional UPA than that in the lens antenna array to achieve the same  array aperture, since the energy focusing capability of the EM lens effectively reduces the   number of antenna elements required in lens array. This may compensate the additional cost of EM lens production and integration in practice. Denote by $\mathbf H_{\upa}(t)\in \mathbb{C}^{M_\upa \times Q_\upa}$ the channel impulse response matrix in the mmWave MIMO with  UPAs. We then have
\begin{align}
\mathbf H_{\upa}(t)=\sum_{l=1}^L \alpha_l \mathbf a_{R,\upa} (\phi_{R,l}) \mathbf a_{T,\upa}^ H(\phi_{T,l}) \delta(t-\tau_l), \label{eq:CIRMatrixUPA}
\end{align}
where $\alpha_l, \tau_l, \phi_{R,l}$ and $\phi_{T,l}$ are defined in \eqref{eq:CIRMatrix}, and $\mathbf a_{R,\upa}$ and $\mathbf a_{T,\upa}$ are the array response vectors corresponding to the UPAs at the receiver and transmitter, respectively, which are given by phase shifting across different antenna elements as \cite{474}
\begin{align}
\mathbf a_{R,\upa}(\phi)&=\begin{matrix}\sqrt{\frac{A_R}{M_\upa}} \Big[1,& e^{j\Phi_2(\phi)},&\cdots & e^{j\Phi_{M_\upa}(\phi)} \Big]\end{matrix}^T, \label{eq:aRUPA}\\
\mathbf a_{T,\upa}(\phi)&=\begin{matrix} \sqrt{\frac{A_T}{Q_\upa}} \Big[1,& e^{j\Phi_2(\phi)}, & \cdots & e^{j\Phi_{Q_\upa}(\phi)} \Big]\end{matrix}^T,\label{eq:aTUPA}
\end{align}
with $\Phi_m$, $m=2,\cdots, M_\upa$ or $2,\cdots, Q_\upa$, denoting the phase shift of the $m$th array element relative to the first antenna. The input-output relationships for the UPA-based wide-band/narrow-band mmWave MIMO communications can be similarly obtained  as in \eqref{eq:rtMIMO} and \eqref{eq:rtMIMONB}, respectively, and are thus omitted for brevity.

 In this paper, we assume that the MIMO channel is perfectly known at the transmitter and receiver for both the proposed lens MIMO and the benchmark UPA-based MIMO systems.

\section{Lens MIMO under Ideal AoAs and AoDs}\label{sec:idealAoAAoD}
To demonstrate the fundamental gains of the lens MIMO based mmWave communication, we first consider an ``ideal'' multi-path propagation environment, where  the spatial frequencies $\{\tphi_{R,l},\tphi_{T,l}\}_{l=1}^L$ corresponding to the AoAs/AoDs of the $L$ paths are all integer multiples  of the spatial frequency resolutions of the receiving/transmitting lenses, i.e., $\{\epsilon_{R,l}, \epsilon_{T,l} \}_{l=1}^L$  defined in \eqref{eq:tphiMIMO} are all zeros. Furthermore, we assume that all the $L$ signal paths have distinct AoAs/AoDs such that $m_{l'}\neq m_l$ and $q_{l'}\neq q_l$, $\forall l'\neq l$. In this case, we show that the multi-path signals in the lens antenna enabled mmWave MIMO system can be perfectly resolved in the spatial domain, thus leading to a new and capacity-achieving spatial multiplexing technique  called OPDM. We also show that with OPDM, the lens antenna based mmWave MIMO system achieves the same (or even better) capacity performance in the narrow-band (wide-band) communications as compared to the conventional UPA based mmWave MIMO, but with dramatically reduced signal processing complexity and RF chain cost.

\subsection{Orthogonal Path Division Multiplexing}\label{sec:OPDM}
In the ``ideal'' AoA/AoD environment as defined above,  the channel impulse response from the transmitting antenna $q$ to receiving antenna $m$ given in \eqref{eq:CIR2} reduces to
\begin{align}\label{eq:CIRIdeal}
h_{mq}(t)=\sum_{l=1}^L \alpha_l \sqrt{A_RA_T}\delta(m-m_l)\delta(q-q_l) \delta(t-\tau_l).
\end{align}
The expression in \eqref{eq:CIRIdeal} implies that the signal transmitted by antenna $q$ will be received at antenna $m$ if and only if there exists a propagation path such that the focusing points corresponding to its AoA and AoD  align exactly with the locations of antenna $m$ and $q$, respectively, i.e., $m=m_l$ and $q=q_l$. Denote by $x_q(t)$ the signal sent by  antenna $q$ of the transmitting lens array, where $q\in \mathcal Q$. The  signal received by antenna $m$ (by ignoring additive noise for the time being) can then be expressed as
\begin{align}
r_m(t)&= \sum_{q\in \mathcal{Q}} h_{mq}(t) \ast x_q(t) \notag \\
&=\begin{cases}
 \sqrt{A_RA_T} \alpha_l x_{q_l}(t-\tau_l) , \ & \text{if } m=m_l \text{ for some } l,\\
0,& \text{otherwise}.\label{eq:ymIdeal}
\end{cases}
\end{align}
Under the assumption of perfect time synchronization at each of the receiving antennas, i.e., $\tau_l$ is known at the receiver and perfectly compensated at antenna $m_l$,  \eqref{eq:ymIdeal} can be equivalently written as
\begin{align}
r_{m_l}=  \sqrt{A_RA_T} \alpha_l x_{q_l}+z_{m_l}, \ l=1,\cdots,L, \label{eq:yml}
\end{align}
where $z_{m_l}$ denotes the AWGN at receiving antenna $m_l$. Therefore, the original  multi-path MIMO channel has been decoupled into $L$ parallel SISO AWGN channels, each corresponding to one of the $L$ multi-paths.  It is worth mentioning that the channel decomposition in \eqref{eq:yml} holds for both the narrow-band and wide-band communications.  This thus enables a new low-complexity and cost-effective way to implement MIMO {\it spatial multiplexing}, by multiplexing $L$ data streams each over one of the $L$ multi-paths independently, which we term as OPDM.

\begin{figure}
\centering
\includegraphics[scale=0.9]{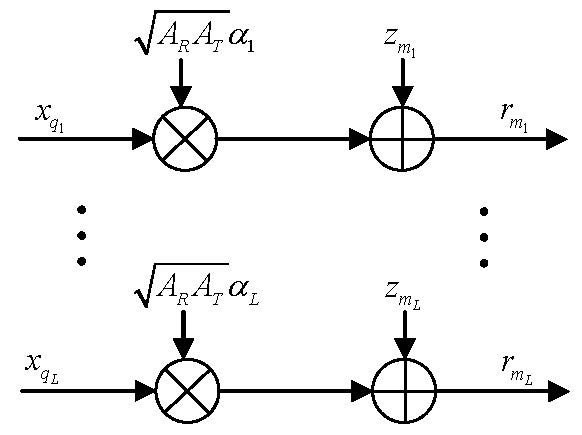}
\caption{Equivalent input-output relationship for OPDM.}\label{F:OPDMEqvSISO}
\end{figure}

A schematic diagram of the equivalent input-output relationship for  OPDM  is shown in Fig.~\ref{F:OPDMEqvSISO}. It is straightforward to show that by applying the standard water-filling (WF) power allocation \cite{56} over each of the $L$ parallel sub-channels with power gains $\{|\alpha_l|^2 A_RA_T\}_{l=1}^L$, the capacity of the mmWave lens MIMO system can be achieved for both narrow-band and wide-band communications.

\subsection{Capacity Comparison}\label{sec:capacityComp}
Next, we provide capacity comparison  by simulations for the proposed lens MIMO versus the conventional UPA-based MIMO in mmWave communications. For the lens MIMO system, we assume that the transmitter and receiver lens apertures are both given by $A_T=A_R=20$, and the effective azimuth lens dimensions are $\tD_T=\tD_R=10$, which corresponds to the number of transmitting/receiving antennas as $M=Q=21$. For fair comparison, the UPA-based MIMO system is assumed to have the same array apertures as the lens MIMO, which thus needs $M_\upa=Q_\upa=80$ transmitting/receiving antennas, as discussed in Section~\ref{sec:benchmark}.
We consider a mmWave channel of $L=3$ paths, which is typical in mmWave communications \cite{486}. We assume a set of ideal AoAs/AoDs with $\tphi_{T,l}=\tphi_{R,l}\in \{0, \pm 0.2\}$. Furthermore, the complex-valued path gains $\{\alpha_l\}_{l=1}^L$ are modeled as $\alpha_l=\sqrt{\beta \kappa_l} e^{j\eta_l}$, $l=1,\cdots, L$ \cite{565}, where $\beta$ denotes the large-scale attenuation including distance-dependent path loss and shadowing, $\kappa_l$ represents the power fractional ratio for the $l$th path, with $\sum_{l=1}^L \kappa_l=1$, and $\eta_l\sim U[0,2\pi]$ denotes the phase shift of the $l$th path. The value of $\beta$ is set based on the generic model $-\beta_{\text{dB}}=c_1+10c_2 \log_{10}(d)+\xi$, where $c_1$ and $c_2$ are the model parameters, $d$ is the communication distance in meters, and $\xi\sim \mathcal{N}(0, \epsilon^2)$ denotes the lognormal shadowing. We assume that the system is operated at the mmWave frequency $f=73$ GHz, for which extensive channel measurements have been performed and the model parameters have been obtained as $c_1=86.6$, $c_2=2.45$, and $\epsilon=8$ dB \cite{565}.  Furthermore, we assume  $d=100$ meters, with which the path loss is $136$ dB, or $\xE[\beta]=-136$ dB, with the expectation taken over the log-normal shadowing.
 In addition, the multi-path power distribution $\{\kappa_l\}_{l=1}^L$ can be modeled as $\kappa_l=\frac{\kappa_l'}{\sum_{k=1}^L \kappa_k'}$, with $\kappa_k'=U_k^{r_\tau-1}10^{-0.1 Z_k}$, where $U_k\sim U[0,1]$ and $Z_k \sim \mathcal{N}(0, \zeta^2)$ are random variables  accounting for the variations in delay and in lognormal shadowing among different paths, respectively~\cite{565}. For mmWave channels at $f=73$ GHz,  $r_\tau$ and $\zeta$ have been obtained as $r_\tau=3$ and $\zeta=4$~\cite{565}. Furthermore, we assume that the total bandwidth is $W=500$ MHz, and the noise power spectrum density is $N_0=-174$ dBm/Hz. Denote by $P$ the total transmission power, the average signal-to-noise ratio (SNR) at each receiving array element  (without the lens applied yet) is then defined as SNR$\triangleq P\xE[\beta]/\sigma^2$. We consider two communication environments characterized by different values of the maximum multi-path excessive delays $T_m$, which correspond to: i) the narrow-band channel with $T_m\ll 1/W$; and ii) the wide-band channel with $T_m=100$~ns.

\begin{figure}
\centering
\includegraphics[scale=0.6]{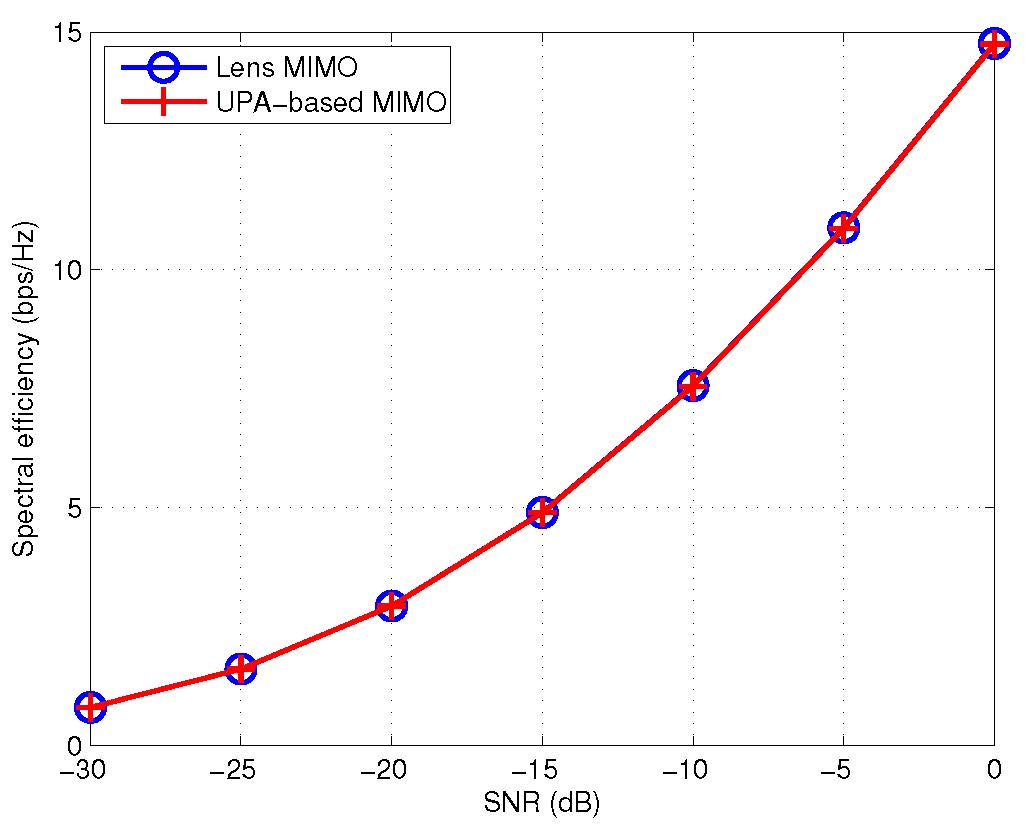}
\caption{Capacity comparison of the lens MIMO using OPDM versus the UPA-based MIMO using eigenmode transmission in narrow-band mmWave communication.}\label{F:RateVsSNRidealAoAAoDNBMain}
\end{figure}

 In Fig.~\ref{F:RateVsSNRidealAoAAoDNBMain}, the average spectrum efficiency is plotted against SNR for both the lens-based  and the UPA-based mmWave MIMO systems  in narrow-band communication,   over $10^4$ random channel realizations. Note that for the UPA-based narrow-band MIMO system, the channel capacity is achieved by the well-known eigenmode transmission with  WF power allocation based on singular value decomposition (SVD) over the MIMO channel matrix \cite{56}.  It is observed from Fig.~\ref{F:RateVsSNRidealAoAAoDNBMain} that under the ideal AoA/AoD environment, the lens MIMO using OPDM achieves almost the same capacity as that by the conventional UPA-based MIMO. However, their required signal processing complexity and hardware cost are rather different, as will be shown in the next subsection.

\begin{figure}
\centering
\includegraphics[scale=0.6]{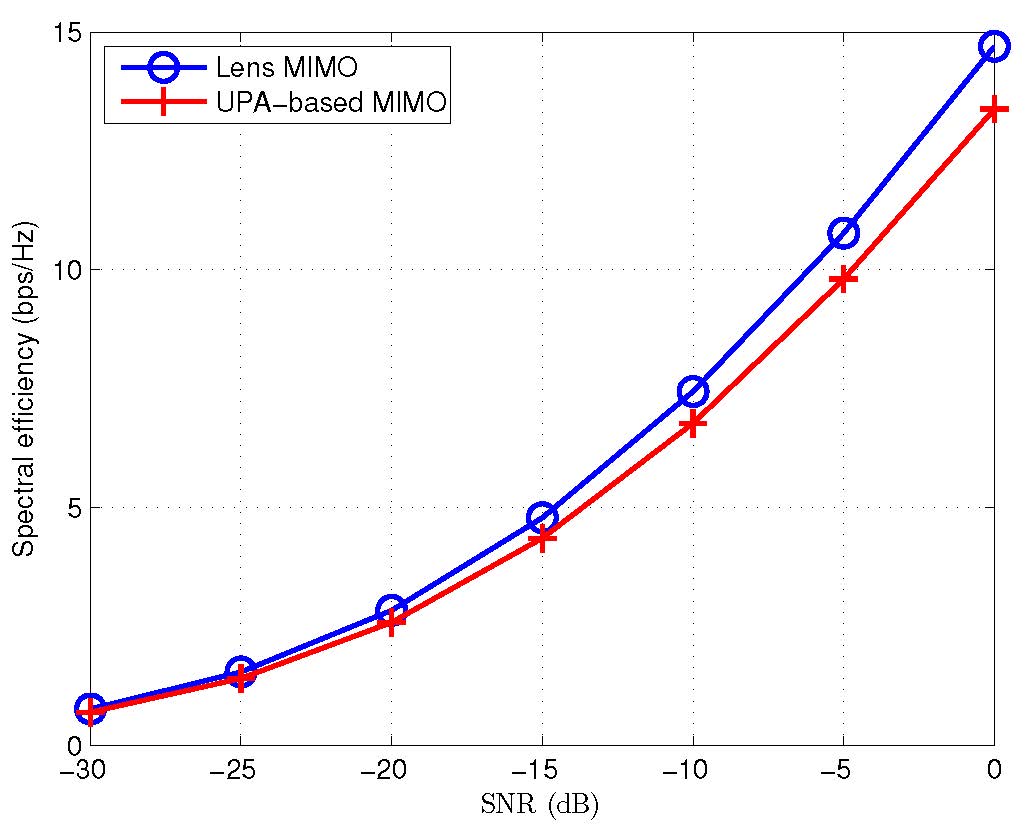}
\caption{Capacity comparison of the lens MIMO using OPDM versus the UPA-based MIMO using MIMO-OFDM in wide-band mmWave communication.}\label{F:RateVsSNRIdealAoAAoDWB}
\end{figure}

  Fig.~\ref{F:RateVsSNRIdealAoAAoDWB} compares the lens MIMO using OPDM versus the UPA-based MIMO using MIMO-OFDM in   wide-band communication. For  MIMO-OFDM, the total bandwidth  is divided into $N=512$ orthogonal sub-bands, and a cyclic prefix (CP) of length $100$ ns is assumed. It is observed in Fig.~\ref{F:RateVsSNRIdealAoAAoDWB} that  for the wide-band communication case, the lens MIMO achieves higher capacity than the UPA-based MIMO, which is mainly due to the time overhead saved for CP transmission.

\begin{table*}
\centering
\caption{Complexity and cost comparison for lens MIMO versus UPA-based MIMO.}\label{table:comparison}
\begin{tabular}{|c|c|c||c|c||c|c|}
\hline
 & \multicolumn{4}{c||}{\bf Signal Processing Complexity}  &  \multicolumn{2}{c|}{\bf Hardware Cost}  \\
\cline{2-7}
%{} & \multicolumn{2}{c}{Precoding/Detection} & \multicolumn{2}{c}{Channel estimation} & \multirow{2}{*}{Antenna} & \multirow{2}{*}{RF chain}\\
 & \multicolumn{2}{c||}{MIMO Processing} & \multicolumn{2}{c||}{Channel Estimation} & \multirow{2}{*}{Antenna} & \multirow{2}{*}{RF chain}\\
%\hline
\cline{2-5}
 & Narrow-band & Wide-band  & Narrow-band & Wide-band & {} & {} \\
\hline
{\bf Lens MIMO} & $O(L)$ & $O(L)$ & $O(L)$ & $O(L)$ & $M+Q$ & $2L$\\
\hline
{\bf UPA-based MIMO} & $O(M_\upa Q_\upa L)$ & $O(M_\upa Q_\upa L N+ (Q_\upa+M_\upa)N\log N)$ & $O(M_\upa Q_\upa)$ & $O(M_\upa Q_\upa N)$  &$M_\upa + Q_\upa$ & $M_\upa + Q_\upa$\\
%{} & Narrow-band & Wide-band & Narrow-band & Wide-band & & \\
%Lens MIMO \\
%Conventional MIMO (without lens)
\hline
\end{tabular}
\end{table*}

\subsection{Complexity and Cost Comparison}
In this subsection, we compare the lens MIMO against the conventional UPA-based MIMO in mmWave communications in terms of signal processing complexity and  hardware cost. The results are  summarized in Table~\ref{table:comparison} and discussed in the following aspects:
\begin{itemize}
\item {\it MIMO processing:} For the lens MIMO based mmWave communication, the capacity for both the narrow-band and wide-band channels is achieved by the simple OPDM scheme, which can be efficiently implemented with signal processing complexity of $O(L)$, with $O(\cdot)$ representing the standard ``big O'' notation.      In contrast, for the UPA-based mmWave MIMO communication, the capacity is achieved by the eigenmode transmission for narrow-band channel and  approached by  MIMO-OFDM for wide-band channel. The signal processing complexity for both schemes mainly arise from determining the eigen-space of the MIMO channel matrices, which has the complexity  $O(M_\upa Q_\upa \min \{M_\upa, Q_\upa\})$ for a generic matrix of size $M_\upa \times Q_\upa$ \cite{563}. For a low-rank $M_\upa \times Q_\upa$ channel matrix of rank $L$, the complexity can be reduced to $O(M_\upa Q_\upa L)$ by exploiting its low-rank property \cite{563}. Thus, the MIMO precoding/detection complexity for the UPA-based MIMO communication is  $O(M_\upa Q_\upa L)$ and $O(M_\upa Q_\upa L N)$  in narrow-band and wide-band communications, respectively, where $N$ denotes the total number of sub-carriers in MIMO-OFDM, which in general requires additional complexity of $O((Q_\upa+M_\upa)N\log N)$   at  the transmitter and receiver for OFDM modulation/demodulation. As $L\ll \min \{M_\upa, Q_\upa\}$ in mmWave communications, the lens MIMO has a significantly lower signal processing complexity than the UPA-based MIMO, especially for the wide-band communication case.
\item {\it Channel estimation:} It follows from \eqref{eq:yml} that the lens MIMO using OPDM only requires estimating $L$ parallel SISO channels for both narrow-band and wide-band communications, which has a complexity $O(L)$. In contrast, the conventional UPA-based MIMO in general requires estimating the MIMO channel of size $M_\upa \times Q_\upa$ for narrow-band communication, and $N$ different MIMO channels each of size  $M_\upa \times Q_\upa$ for  wide-band communication using MIMO-OFDM.\footnote{Note that by exploiting the channel sparsity in mmWave communications with small $L$, the channel estimation in UPA-based MIMO can be implemented with   lower complexity  via jointly estimating the multi-path  parameters $\{\alpha_l, \phi_{R,l}, \phi_{T,l}, \tau_l\}_{l=1}^L$, which, however, requires  more sophisticated techniques as in \cite{579}.}
%\item {\it Antenna cost:} As mentioned in Section~\ref{sec:modelLensMIMO}, thanks to the energy focusing provided by the EM lenses, the lens antenna array in general requires less antenna elements than the conventional array for the same effective array aperture, i.e., $M< M_\upa$ and $Q< Q_\upa$. On the other hand, it requires the deployment of the additional EM lenses each at the transmitter and the receiver, which are in fact inexpensive with modern lens fabrication technologies \cite{554,559}.
\item {\it Hardware cost:} The  hardware cost for mmWave MIMO communications mainly depends on the  required number of transmitting/receiving RF chains, which are composed of  mixers, amplifiers, D/A or A/D converters, etc.
    For the lens MIMO system, it follows from \eqref{eq:yml} that only  $L$ receiving/transmitting antennas located on the focusing points of the $L$ multi-paths need to be selected to operate at one time; whereas all the remaining antennas can be deactivated. This thus helps to significantly reduce the number of RF chains required as compared to the conventional UPA-based MIMO, as shown in Table~\ref{table:comparison} in detail.
\end{itemize}

\section{Lens MIMO under Arbitrary AoAs/AoDs}\label{sec:arbAoAAoD}
\begin{figure}
\centering
\includegraphics[scale=0.6]{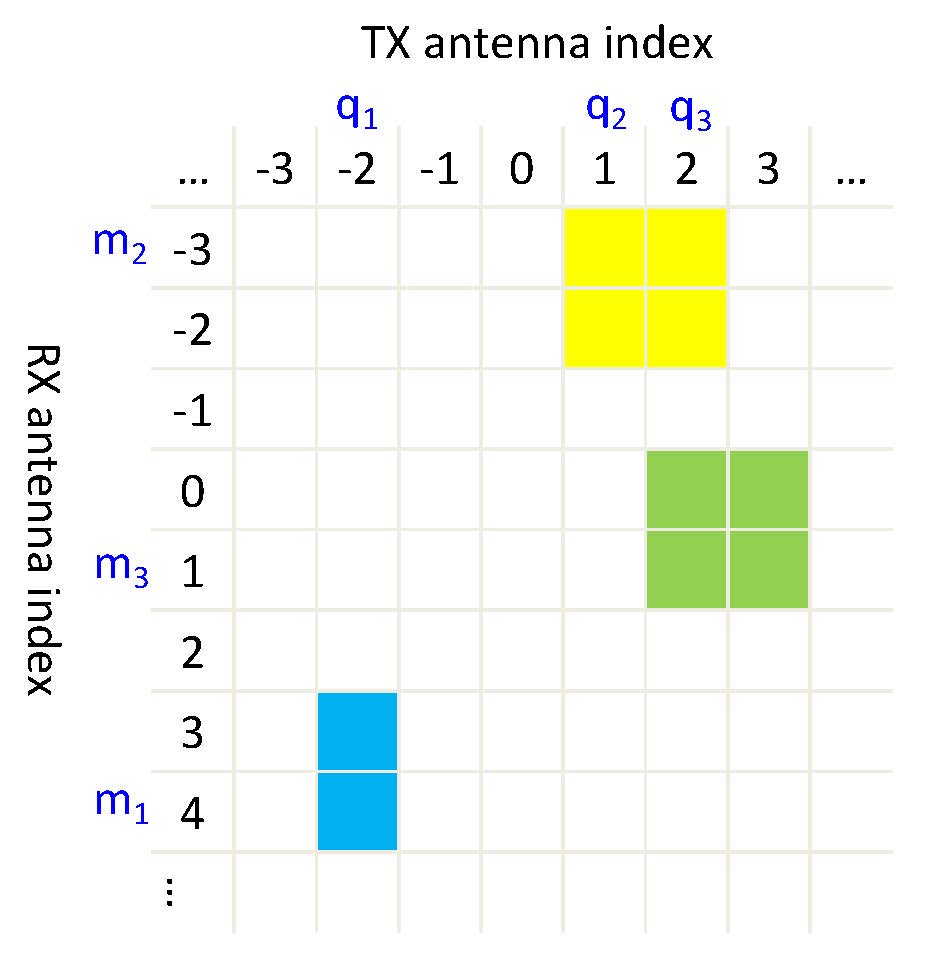}
\caption{An illustration of the sparsity for a lens MIMO channel with $L=3$ paths. $\tD_T=\tD_R=10$. $\tphi_{R,l}\in \{0.36, -0.27, 0.08\}$, $\tphi_{T,l}\in \{-0.2, 0.12, 0.24\}$. $\Delta=1$. $\mathcal M_1=\{3,4\}, \mathcal M_2=\{-3, -2\}, \mathcal M_3=\{0,1\}, \mathcal Q_1=\{-2\}, \mathcal Q_2=\{1, 2\}$, and $\mathcal Q_3=\{2, 3\}$. Note that $\mathcal Q_2 \cap \mathcal Q_3 \neq \emptyset$ due to the small AoD separation between path 2 and path 3.}\label{F:SparseChannel}
\end{figure}

In this section, we study the mmWave lens MIMO  in the  general channel  with arbitrary AoAs/AoDs, i.e., the spatial frequencies $\{\tphi_{R,l}, \tphi_{T,l}\}_{l=1}^L$ are not necessarily integer multiples of the spatial frequency resolutions of the receiving/transmitting lens arrays.
In this case, the power for each multi-path signal in general  spreads across the entire antenna array with decaying power levels from the antenna closest to the corresponding  focusing point. Let $\Delta>0$ be a positive integer with which it can be practically approximated that $|\sinc(x)|^2\approx 0$, $\forall |x|\geq \Delta$.\footnote{For practical applications, $\Delta=1$ is a reasonable choice, since $|\sinc(x)|^2\leq 0.047$, $\forall |x|\geq 1$.} It then follows from \eqref{eq:arrayResponseRX} and \eqref{eq:arrayResponseTX} that the receive/transmit  array responses for the $l$th path are negligible at those antennas with a distance greater than $\Delta$ from the focusing point (see Fig.~\ref{F:ArrayResponse}), i.e.,
\begin{align}
a_{R,m}(\phi_{R,l})&=\sqrt{A_R} \sinc(m-\tD_R\tphi_{R,l}) \approx 0,  \quad \forall  m \notin \mathcal{M}_l, \\
a_{T,q}(\phi_{T,l})&=\sqrt{A_T}\sinc(q-\tD_T\tphi_{T,l}) \approx 0,   \quad \forall  q\notin \mathcal{Q}_l,
\end{align}
where $\mathcal{M}_l$ and $\mathcal{Q}_l$ are referred to as the {\it supporting} receiving/transmitting antenna subsets for the $l$th path, which are defined as
\begin{align}
\mathcal{M}_l & \triangleq \left\{m\in \mathcal{M} : |m-\tD_R \tphi_{R,l}|<\Delta\right\}, \label{eq:Ml}\\
\mathcal{Q}_l & \triangleq \left\{q\in \mathcal{Q} : |q-\tD_T \tphi_{T,l}|<\Delta \right \}, \ l=1,\cdots, L. \label{eq:Ql}
\end{align}
%In the following, $\mathcal M_l$ and $\mathcal Q_l$ are sometimes referred to as the receiving and transmitting antenna {\it supports} for path $l$, respectively.
Consequently, the $(m,q)$-th element of the channel impulse response matrix $\mathbf H(t)$ in \eqref{eq:CIRMatrix} has practically non-negligible power if and only if there exists at least one signal path $l$ such that $m\in \mathcal M_l$ {\it and} $q\in \mathcal Q_l$. Since $L\ll \min \{M,Q\}$ due to the multi-path sparsity in mmWave systems, it follows that $\mathbf H(t)$ is in practice a (nearly) sparse matrix with block sparsity structure, where each non-zero block corresponds to one of the $L$ multi-paths and has approximately $2\Delta \times 2\Delta$ entries around the element $(m_l, q_l)$, as illustrated in Fig.~\ref{F:SparseChannel}. %Such a channel sparsity structure can be exploited to design low-complexity and low-cost transceivers for both narrow-band and wide-band mmWave communications, as will be pursued in the following subsections.
Note that depending on the AoA/AoD values, $\{\mathcal M_l\}_{l=1}^L$ (or $\{\mathcal Q_l\}_{l=1}^L$) may have non-empty intersection for different paths, i.e., certain antenna elements may receive/transmit non-negligible   power from/to more than one signal paths, as illustrated by $\mathcal Q_2$ and $\mathcal Q_3$ in Fig.~\ref{F:SparseChannel}.

Let $\mathcal M_S = \bigcup_{l=1}^L \mathcal{M}_l$ and $\mathcal Q_S = \bigcup_{l=1}^L \mathcal {Q}_l$ be the supporting receiving/transmitting antenna subsets associated with all the $L$ paths, and $\mathbf H_S(t)\in \mathbb{C}^{|\mathcal M_S|\times |\mathcal Q_S|}$ be the  sub-matrix of the channel impulse response $\mathbf H(t)$ corresponding to the receiving antennas in $\mathcal M_S$ and transmitting antennas in $\mathcal Q_S$. By deactivating those antennas with negligible channel powers, the input-output relationship in \eqref{eq:rtMIMO} then reduces to
\begin{align}
\mathbf r_{\mathcal M_S}(t)&= \mathbf H_S(t) \ast \mathbf x_{\mathcal Q_S}(t)+\mathbf z_{\mathcal M_S}(t)\\
&=\sum_{l=1}^L \alpha_l \mathbf a_{R, \mathcal M_S}(\phi_{R,l}) \mathbf a_{T,\mathcal Q_S}^H (\phi_{T,l})\mathbf x_{\mathcal Q_S}(t-\tau_l)+\mathbf z_{\mathcal M_S}(t),\label{eq:rMS}
\end{align}
where $\mathbf r_{\mathcal M_S}, \mathbf a_{R,\mathcal M_S}, \mathbf z_{\mathcal M_S}\in \mathbb{C}^{|\mathcal M_S|\times 1}$ respectively denote the sub-vectors of $\mathbf r, \mathbf a_R$ and $\mathbf z$ in \eqref{eq:rtMIMO} corresponding to the receiving antennas in $\mathcal M_S$; and $\mathbf a_{T, \mathcal Q_S}, \mathbf x_{\mathcal Q_S}\in \mathbb{C}^{|\mathcal Q_S|\times 1}$ denote the sub-vectors of $\mathbf a_T$ and $\mathbf x$ corresponding to the transmitting antennas in $\mathcal Q_S$, respectively. %Note that since $\mathf a_{R,\mathcal M_S}(\phi_{R,l})$
\begin{remark}
It follows from \eqref{eq:rMS} that for mmWave lens MIMO system with arbitrary AoAs/AoDs,  only $|\mathcal M_S|\ll M$ receiving and $|\mathcal Q_S|\ll Q$ transmitting RF chains are generally needed to achieve the near-optimal performance of the full-MIMO system with all $M+Q$ antennas/RF chains in use. Furthermore, since $|\mathcal M_S|\leq \sum_{l=1}^L |\mathcal M_l|\approx 2\Delta L$, and $|\mathcal Q_S|\leq \sum_{l=1}^L |\mathcal Q_l| \approx 2\Delta L$, the total number of RF chains required only depends on the number of multi-paths $L$, instead of the actually deployed antennas $M$ and $Q$.
\end{remark}

%{\color{red}
\subsection{Transceiver Design Based on PDM }\label{sec:PDM}
In this subsection, by exploiting the reduced-size channel matrix in \eqref{eq:rMS}, we propose a low-complexity transceiver design based on PDM (instead of OPDM due to arbitrary AoAs/AoDs),  which is applicable for both narrow-band and wide-band mmWave communications. With PDM, $L$ independent data streams are transmitted in general, each through  one of the $L$ multi-paths by transmit beamforming/precoding. Specifically, the discrete-time equivalent of the transmitted signal $\mathbf x_{\mathcal Q_S}(t)$  can be expressed as
\begin{align}
\mathbf x_{\mathcal Q_S}[n]=\sum_{l=1}^L \sqrt{\frac{p_l}{A_T}}\mathbf a_{T, \mathcal Q_S}(\phi_{T,l})s_l[n], \label{eq:xQS}
\end{align}
where $n$ denotes the symbol index, $s_l[n]\sim \mathcal{CN}(0,1)$ represents the i.i.d. CSCG distributed information-bearing symbols for data stream $l$, with transmit power $p_l$; and $\mathbf a_{T, \mathcal Q_S}(\phi_{T,l})/\sqrt{A_T}$ denotes the unit-norm \emph{per-path} MRT beamforming vector towards the AoD $\phi_{T,l}$ of path $l$. Note that we have used the identity $\|\mathbf a_{T, \mathcal Q_S}(\phi_{T,l})\|^2\approx \|\mathbf a_{T}(\phi_{T,l})\|^2=A_T$, $\forall l$.  At the receiver side, the low-complexity per-stream based detection is used, where a receiving beamforming vector $\mathbf v_{l}\in \mathbb{C}^{|\mathcal M_S|\times 1}$ with $\|\mathbf v_l\|=1$ is applied over the receiving antennas in $\mathcal M_S$ for detecting   $s_l[n]$. Thus, we have
 \begin{align}
 \hat s_l[n] = \mathbf v_l^H \mathbf r_{\mathcal M_S}[n], \ l=1,\cdots, L, \label{eq:hatsl2}
 \end{align}
 where $\mathbf r_{\mathcal M_S}[n]$ is the discrete-time equivalent of the received signal $\mathbf r_{\mathcal M_S}(t)$ shown in \eqref{eq:rMS}. %A schematic diagram of the PDM scheme is shown in Fig.~\ref{}.

 Next, we analyze the performance of the above proposed PDM scheme for wide-band communications. The analysis for the special case of narrow-band communications can be obtained similarly and is thus omitted for brevity. For simplicity, we assume that the multi-path delays can be approximated as  integer multiples of the symbol interval $T_s$, i.e., $\tau_l=n_l T_s$ for some integer $n_l$, $\forall l$. For notational conciseness, let $\mathbf a_{T,l}\triangleq \mathbf a_{T, \mathcal Q_S}(\phi_{T,l})$ and $\mathbf a_{R,l}\triangleq \mathbf a_{R, \mathcal M_S}(\phi_{R,l})$, $\forall l$. Based on \eqref{eq:rMS} and \eqref{eq:xQS}, the discrete-time equivalent received signal $\mathbf r_{\mathcal M_S}[n]$ can be expressed as
 \begin{align}
 \mathbf r_{\mathcal M_S}& [n] = \sum_{k=1}^L \alpha_k \mathbf a_{R,k} \mathbf a_{T,k}^H\mathbf x_{\mathcal Q_S}[n-n_k] + \mathbf z_{\mathcal M_S}[n]\\
=&\underbrace{\sqrt{p_l A_T} \alpha_l \mathbf a_{R,l} s_l[n-n_l]}_{\text{desired signal}} + \underbrace{\sum_{k\neq l}^L \sqrt{\frac{p_l}{A_T}} \alpha_k \mathbf a_{R,k} \mathbf a_{T,k}^H \mathbf a_{T,l} s_l[n-n_k]}_{\text{ISI}} \notag \\
&+ \underbrace{\sum_{l'\neq l}^L \sum_{k=1}^L \sqrt{\frac{p_{l'}}{A_T}} \alpha_k \mathbf a_{R,k}\mathbf a_{T,k}^H \mathbf a_{T,l'}s_{l'}[n-n_k]}_{\text{inter-stream interference}} + \mathbf z_{\mathcal M_S}[n]. \label{eq:rMS4}
 \end{align}
Note that in  \eqref{eq:rMS4}, we have decomposed the received signal $\mathbf r_{\mathcal M_S}[n]$ from the perspective of the $l$th data stream, which includes the desired signal component propagated via the $l$th path with symbol delay $n_l$, the ISI from the same data stream received via all other $L-1$ paths with different delays, and the inter-stream interference from the other $L-1$ data streams. By  applying the receiver beamforming in \eqref{eq:hatsl2} and treating the ISI and the inter-stream interference both as noise, the effective SNR for the $l$th data stream can be expressed as \eqref{eq:gammal} shown at the top of the next page. 
\begin{figure*}
\begin{align}
\gamma_l=\frac{p_l A_T |\alpha_l|^2 |\mathbf v_l^H \mathbf a_{R,l}|^2}{\sum_{k\neq l}^L \frac{p_l}{A_T} |\alpha_k|^2 |\mathbf v_l^H \mathbf a_{R,k}|^2 |\mathbf a_{T,k}^H \mathbf a_{T,l}|^2+\sum_{l'\neq l}^L \sum_{k=1}^L \frac{p_{l'}}{A_T}|\alpha_k|^2 |\mathbf v_l^H \mathbf a_{R,k}|^2 |\mathbf a_{T,k}^H \mathbf a_{T,l'}|^2 + \sigma^2}, \forall l. \label{eq:gammal}
\end{align}
\end{figure*}
The achievable sum-rate is then given by $R=\sum_{l=1}^L \log_2(1+\gamma_l)$. In the following, the two commonly used receiver beamforming schemes, i.e., MRC and MMSE beamforming, are studied to gain  insights on the proposed PDM transmission scheme.

\subsubsection{MRC Receive Beamforming}
With MRC, the receiver beamforming vector $\mathbf v_l$ for data stream $l$ is set to maximize the desired signal power from the $l$th path, i.e., $\mathbf v_l^{\text{MRC}} = \mathbf a_{R,l}/\sqrt{A_R}$, $\forall l$.  By substituting $\mathbf v_l^{\text{MRC}}$ into \eqref{eq:gammal}, the SNR can be expressed as \eqref{eq:SNRMIMOWB} shown at the top of the next page.
\begin{figure*}
\vspace{-3ex}
\begin{align}
\gamma_l^{\text{MRC}} = \frac{p_l  |\alpha_l|^2 A_RA_T}{\sum_{k\neq l}^L \frac{p_l}{A_RA_T} |\alpha_k|^2 \left|\mathbf a_{R,l}^H \mathbf a_{R,k}\right|^2 \left|\mathbf a_{T,k}^H \mathbf a_{T,l} \right|^2
+ \sum_{l'\neq l}^L \sum_{k=1}^L \frac{p_{l'}}{A_RA_T} |\alpha_k|^2 \left|\mathbf a_{R,l}^H \mathbf a_{R,k} \right|^2 \left|\mathbf a_{T,k}^H\mathbf a_{T,l'}\right|^2 + \sigma^2}.\label{eq:SNRMIMOWB}
\end{align}
\hrule
\end{figure*}
Note that we have used the identity $\|\mathbf a_{R,l}\|^2\approx A_R$, $\forall l$. For two different paths $l'\neq l$, define the transmitter and receiver side {\it inter-path contamination} (IPC) coefficients   as
\begin{align}
\rho_T^{ll'}\triangleq \frac{\left|\mathbf a_{T,l}^H \mathbf a_{T,l'}\right|^2}{A_T^2}< 1, \quad %= \frac{\left|(\mathbf a_T^{l})^H \mathbf a_T^{l'} \right|^2}{\left\| \mathbf a_T^{l} \right\|^2 \left\| \mathbf a_T^{l'}\right\|^2}\\
\rho_R^{ll'}\triangleq \frac{\left|\mathbf a_{R,l}^H \mathbf a_{R,l'}\right|^2}{A_R^2}< 1. %= \frac{\left|(\mathbf a_R^{l})^H \mathbf a_R^{l'} \right|^2}{\left\| \mathbf a_R^{l} \right\|^2 \left\| \mathbf a_R^{l'}\right\|^2}.
\end{align}
The SNR in \eqref{eq:SNRMIMOWB} can then be simplified as
{\small
\begin{align}
\hspace{-4ex} \gamma_l^{\text{MRC}} &= \frac{p_l |\alpha_l|^2}{\sum_{k\neq l}^L p_l |\alpha_k|^2 \rho_R^{lk} \rho_T^{kl} + \sum_{l'\neq l}^L\sum_{k=1}^L p_{l'} |\alpha_k|^2 \rho_R^{lk} \rho_T^{kl'}+\frac{\sigma^2}{A_RA_T}}\label{eq:SNRMIMOWB2} \\
&\approx  \frac{p_l |\alpha_l|^2}{\sum_{k\neq l}^L p_l |\alpha_k|^2 \rho_R^{lk} \rho_T^{kl} + \sum_{l'\neq l}^L p_{l'} \left(|\alpha_{l'}|^2 \rho_R^{ll'} + |\alpha_{l}|^2 \rho_T^{ll'}\right)  +\frac{\sigma^2}{A_RA_T}}, \label{eq:SNRMIMOWB3}
\end{align}}
where the approximation in \eqref{eq:SNRMIMOWB3} is obtained by keeping only the two dominating inter-stream interference terms in  \eqref{eq:SNRMIMOWB2} with either $k=l'$ or $k=l$.

\begin{figure}
\centering
\includegraphics[scale=0.6]{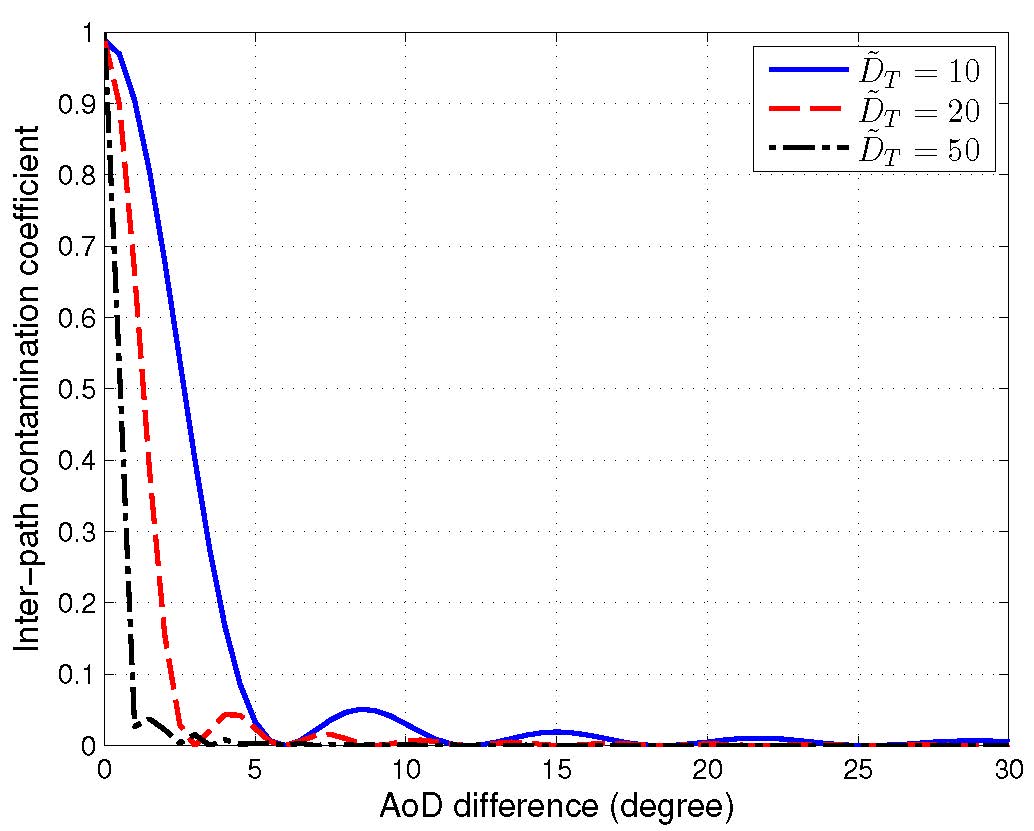}
\caption{Transmitter-side inter-path contamination coefficient versus the AoD difference $|\phi_{T,l}-\phi_{T,l'}|$ with different $\tD_T$   in lens MIMO systems.}\label{F:IPCversAoADifference}
\end{figure}

It is observed from \eqref{eq:SNRMIMOWB3} that for wide-band mmWave lens MIMO systems using PDM and the simple MRC receiver beamforming, the ISI is {\it double} attenuated as can be seen from   the IPC coefficients  $\rho_T^{kl}$ and $\rho_R^{lk}$ at both the transmitter and the receiver sides, and the inter-stream interference is attenuated through either transmitter-side IPC coefficient $\rho_T^{ll'}$ or receiver-side IPC coefficient $\rho_R^{ll'}$. Based on \eqref{eq:arrayResponseTX}, we have
\begin{align}
\rho_T^{ll'} &= \frac{1}{A_T^2}\left| \sum_{q\in \mathcal Q_S} a_{T,q}^*(\phi_{T,l})a_{T,q}(\phi_{T,l'}) \right|^2 \\
&=\left| \sum_{q\in \mathcal Q_S} \sinc(q-\tD_T \tphi_{T,l})\sinc(q-\tD_T \tphi_{T,l'})\right|^2,
\end{align}
which vanishes to zero for sufficiently separated AoDs such that $|\tphi_{T,l}-\tphi_{T,l'}|> 2\Delta/\tD_T$, or equivalently $\mathcal Q_l\cap \mathcal Q_{l'}=\emptyset$. Similarly this holds for the receiver side IPC coefficient $\rho_R^{ll'}$. In Fig.~\ref{F:IPCversAoADifference}, the IPC coefficient $\rho_T^{ll'}$ is plotted against the AoD difference $|\phi_{T,l}-\phi_{T,l'}|$ for different AoD resolutions provided by the transmitter lens array, which verifies that the IPC vanishes asymptotically with large AoD separations and/or high  AoD resolutions.

In the favorable propagation environment  with both sufficiently separated AoAs and AoDs such that $\rho_R^{ll'}\approx 0$ and $\rho_T^{ll'}\approx 0$, $\forall l'\neq l$, both the ISI and the inter-stream interference in \eqref{eq:SNRMIMOWB2} vanish. As a result, the SNR for the $l$th data stream reduces to $\gamma_l=p_l|\alpha_l|^2 A_R A_T /\sigma^2$, $\forall l$, which is identical to that achieved by the OPDM   in the ideal AoAs/AoDs case shown in Fig.~\ref{F:OPDMEqvSISO}. In this case,   PDM with simple MRC receive beamforming achieves the channel capacity for both narrow-band and wide-band mmWave communications.

\subsubsection{MMSE Receive Beamforming}
In the general case where the transmitter- and/or receiver-side IPC coefficients are non-zero due to the limited AoA/AoD separations and/or insufficient AoA/AoD resolutions provided by the lens arrays, the PDM scheme suffers from both ISI and inter-stream interference, which needs to be further mitigated. One simple interference mitigation scheme is via MMSE   beamforming at the receiver, for which the beamforming vector $\mathbf v_l$ in \eqref{eq:hatsl2} for the $l$th data stream is set as \cite{71}
\begin{align}
\mathbf v_l^{\text{MMSE}}=\frac{\mathbf C_l^{-1} \mathbf a_{R,l}}{\|\mathbf C_l^{-1} \mathbf a_{R,l}\|}, \ l=1,\cdots, L,
\end{align}
where $\mathbf C_l$ is the covariance matrix of the effective noise vector. Based on \eqref{eq:rMS4}, $\mathbf C_l$ can be obtained as \eqref{eq:Cl} shown at the top of the next page,
\begin{figure*}
\begin{align}
\mathbf C_l&=\sum_{k\neq l}^L \frac{p_l}{A_T}|\alpha_k|^2 |\mathbf a_{T,k}^H \mathbf a_{T,l}|^2 \mathbf a_{R,k}\mathbf a_{R,k}^H
+\sum_{l'\neq l}^L \sum_{k=1}^L \frac{p_{l'}}{A_T}|\alpha_k|^2 |\mathbf a_{T,k}^H \mathbf a_{T,l'}|^2 \mathbf a_{R,k}\mathbf a_{R,k}^H + \sigma^2 \mathbf I\\
&=\sum_{k\neq l}^L p_lA_TA_R|\alpha_k|^2 \rho_{T}^{kl} \tilde{\mathbf a}_{R,k}\tilde{\mathbf a}_{R,k}^H
+\sum_{l'\neq l}^L \sum_{k=1}^L p_{l'}A_TA_R|\alpha_k|^2 \rho_{T}^{kl'} \tilde{\mathbf a}_{R,k}\tilde{\mathbf a}_{R,k}^H + \sigma^2 \mathbf I,  \forall l, \label{eq:Cl}
\end{align}
\hrule
\end{figure*}
where $\tilde{\mathbf a}_{R,k}\triangleq \mathbf a_{R,k}/\sqrt{A_R}$, $k=1,\cdots, L$.
The corresponding SNR can be obtained as
\begin{align}
\gamma_l^{\text{MMSE}}=&|\alpha_l|^2p_lA_T\mathbf a_{R,l}^H \mathbf C_l^{-1} \mathbf a_{R,l}\\
=&|\alpha_l|^2p_l \tilde{\mathbf a}_{R,l}^H\Bigg( \sum_{k\neq l}^L  p_l |\alpha_k|^2 \rho_{T}^{kl} \tilde{\mathbf a}_{R,k}\tilde{\mathbf a}_{R,k}^H \notag \\
&+ \sum_{l'\neq l}^L \sum_{k=1}^L p_{l'}|\alpha_k|^2 \rho_{T}^{kl'} \tilde{\mathbf a}_{R,k}\tilde{\mathbf a}_{R,k}^H + \frac{\sigma^2}{A_TA_R} \mathbf I\Bigg)^{-1} \tilde{\mathbf a}_{R,l} \label{eq:SNRMMSE}\\
\geq & \gamma_l^{\text{MRC}}, \ l=1,\cdots, L, \label{eq:greater}
\end{align}
where \eqref{eq:greater} follows from the fact that $\mathbf x^H \mathbf A^{-1} \mathbf x \geq 1/(\mathbf x^H \mathbf A \mathbf x)$, $\forall \|\mathbf x\|=1$. The inequality in  \eqref{eq:greater} shows   that MMSE   beamforming in general achieves better performance than MRC, since it strikes a balance between maximizing the desired signal power and minimizing the interference. In the favorable scenario with both sufficiently separated AoAs and AoDs such that $\rho_T^{ll'}\approx 0$ and $\rho_R^{ll'}\approx 0$, $\forall l'\neq l$, it can be shown that the MMSE and MRC receive  beamforming vectors are identical.

\subsection{Path Grouping}
As can be seen from \eqref{eq:SNRMIMOWB2} and \eqref{eq:SNRMMSE}, the performance of the PDM scheme with  MRC or MMSE receive beamforming depends on the ISI and inter-stream interference power via the IPC coefficients $\rho_T^{ll'}$ and $\rho_R^{ll'}$, $\forall l'\neq l$. In this subsection, the PDM scheme is further improved by applying the technique of {\it path grouping}, by which the paths that are significantly interfered with each other are grouped and jointly processed. It is shown that the PDM with path-grouping  achieves the channel capacity for both narrow-band and wide-band lens MIMO systems, provided that {\it either} the AoAs or AoDs (not necessarily both) are sufficiently separated.

\subsubsection{Sufficiently Separated AoAs}\label{sec:sufficientAoAs}
\begin{figure}
\centering
\includegraphics[scale=1]{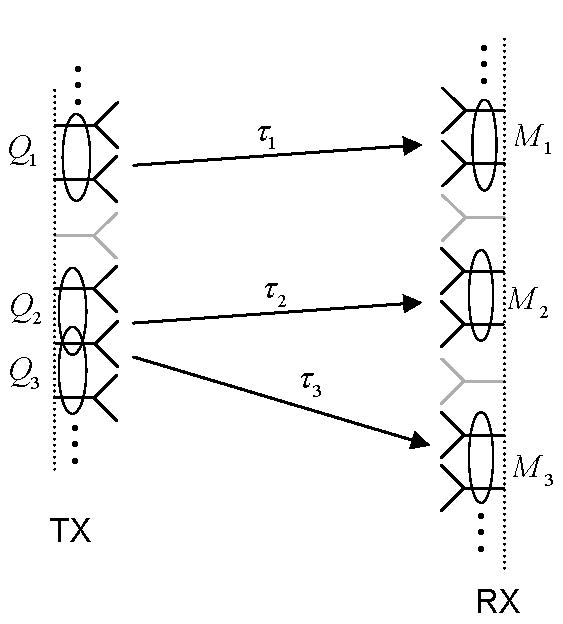}
\caption{An illustration of the effective channel in mmWave lens MIMO system with sufficiently separated AoAs. The path delays are labeled for each link. Gray antennas represent those with negligible power  and hence can be deactivated. Note that path $2$ and path $3$  are grouped since they have similar AoDs at Tx (but different AoAs at Rx).}\label{F:ModelWellSeparatedAoAs}
\end{figure}

We first consider the case with sufficiently separated AoAs for all paths such that $|\tphi_{R,l}-\tphi_{R,l'}|>\frac{2\Delta}{\tD_R}$, $\forall l'\neq l$, but  with  possibly close AoDs for certain paths. This may correspond to the uplink communications where the receiving lens antenna array equipped at the base station has a large azimuth dimension ($\tD_R\gg 1$) and hence provides   accurate AoA resolution; whereas the transmitting lens array at the mobile terminal can only provide moderate AoD resolution. In this case, it follows from \eqref{eq:Ml} that $\mathcal M_l \cap \mathcal M_{l'}=\emptyset$, $\forall l'\neq l$, i.e.,  $\left\{\mathcal M_l\right\}_{l=1}^L$ form a disjoint partition for the supporting receiving antenna subset $\mathcal M_S$. As a result, \eqref{eq:rMS} can be decomposed into
%\begin{equation}
\begin{align}
\mathbf r_{\mathcal M_l}(t)= \alpha_l \mathbf a_{R,\mathcal M_l}(\phi_{R,l})\mathbf a_{T,\mathcal Q_S}^H (\phi_{T,l}) & \mathbf x_{\mathcal Q_S}(t-\tau_l)+ \mathbf z_{\mathcal M_l}(t),\notag \\
&   l=1,\cdots, L,\label{eq:rMl}
\end{align}
%\end{equation}
where $\mathbf r_{\mathcal M_l}, \mathbf a_{R, \mathcal{M}_l}, \mathbf z_{\mathcal M_l}\in \mathbb{C}^{| \mathcal M_l| \times 1}$ are respectively the sub-vectors of  $\mathbf r_{\mathcal M_S}, \mathbf a_{R, \mathcal M_S}$ and $\mathbf z_{\mathcal M_S}$ in \eqref{eq:rMS} corresponding to the receiving antennas in $\mathcal M_l$. \eqref{eq:rMl} shows that each receiving antenna only receives the signals via one of the multi-paths, thanks to the sufficient AoA separations such that the signals from different multi-paths are focused at non-overlapping receiving antenna subsets. However, the signal transmitted by certain transmitting antennas may propagate via more than one paths due to the possible overlapping of the supporting transmitting antenna subsets for different paths. Such a phenomenon is  illustrated in Fig.~\ref{F:ModelWellSeparatedAoAs}.

Due to the single  path received by each receiving antenna, the path delay $\tau_l$ can be compensated by the antennas in $\mathcal M_l$. As a result, \eqref{eq:rMl} is equivalent to
\begin{align}
%\mathbf r_{\mathcal M_l}=\alpha_l \mathbf a_{R,\mathcal M_l}(\phi_{R,l})\mathbf a_{T,\mathcal Q_S}^H (\phi_{T,l})\mathbf x_{\mathcal Q_S}+ \mathbf z_{\mathcal M_l}, l=1,\cdots, L.
\underbrace{\left[\begin{matrix}
\mathbf r_{\mathcal M_1} \\
\vdots \\ \mathbf r_{\mathcal M_L}
\end{matrix}\right]}_{\mathbf r_{\mathcal M_S}}=
\underbrace{\left[
\begin{matrix}
\alpha_1 \mathbf a_{R,\mathcal M_1}(\phi_{R,1})\mathbf a_{T,\mathcal Q_S}^H (\phi_{T,1})\\
\vdots
\\
\alpha_L \mathbf a_{R,\mathcal M_L}(\phi_{R,L})\mathbf a_{T,\mathcal Q_S}^H (\phi_{T,L})
\end{matrix}
\right]}_{\mathbf H_S} \mathbf x_{\mathcal Q_S} +
\underbrace{\left[
\begin{matrix}
\mathbf z_{\mathcal M_1} \\
\vdots \\ \mathbf z_{\mathcal M_L}
\end{matrix}
\right]}_{\mathbf z_{\mathcal M_S}}.\label{eq:rMS2}
\end{align}
In other words, with sufficiently separated AoAs, the original multi-path channel in \eqref{eq:rMS} is essentially equivalent to a simple MIMO AWGN channel given in \eqref{eq:rMS2}, regardless of narrow-band or wide-band communications.\footnote{Recall from Section~\ref{sec:PDM} that with {\it either} sufficiently separated AoAs or AoDs, i.e., $\rho_{R}^{ll'}\approx 0$ or $\rho_{T}^{ll'}\approx 0$, $\forall l'\neq l$, the ISI can be completely eliminated by PDM.} The channel capacity of \eqref{eq:rMS2} is known to be achieved by the  eigenmode transmission with WF power allocation based on the MIMO channel matrix $\mathbf H_S$. However, a closer look at $\mathbf H_S$ reveals that it is still a sparse matrix due to the sparsity of the transmitting response vectors $\mathbf a_{T,\mathcal Q_S}(\phi_{T,l})$, $\forall l$, which can be further exploited to reduce the complexity for achieving the capacity of the MIMO channel in \eqref{eq:rMS2}.

Recall that the transmitting array response vector $\mathbf a_{T,\mathcal Q_S}(\phi_{T,l})$ has essentially non-zero entries only for those transmitting antennas in the subset $\mathcal Q_l\subseteq \mathcal Q_S$. The main idea for the proposed design is  called {\it AoD-based path grouping}, by which the $L$ paths are partitioned into $G\leq L$ groups such that paths $l$ and $l'$ belong to the same group if the transmitter-side IPC coefficient $\rho_{T}^{ll'}>0$, or equivalently if $\mathcal Q_l \cap \mathcal Q_{l'} \neq \emptyset$. Denote by $\mathcal L_g\subseteq \{1,\cdots, L\}$ the path indices in the $g$th group, $g=1,\cdots, G$. For instance, for the system shown in Fig.~\ref{F:ModelWellSeparatedAoAs}, we have $G=2$ and $\mathcal L_1=\{1\}$ and $\mathcal L_2=\{2,3\}$.
   In addition, denote by $\mathcal{\bar Q}_{g} \triangleq \bigcup_{l\in \mathcal L_g} \mathcal Q_l$ and $\bar {\mathcal M}_g\triangleq \bigcup_{l\in \mathcal L_g} \mathcal M_l$, $g=1,\cdots, G$, the supporting transmitting and receiving antenna subsets for all paths in group $g$, respectively. By construction, $\{\mathcal{\bar Q}_{g} \}_{g=1}^G$ and $\{\mathcal{\bar M}_{g} \}_{g=1}^G$ form  disjoint partitions for the supporting transmitting antenna subsets $\mathcal Q_S$ and $\mathcal M_S$, respectively. Therefore, the input-output relationship in \eqref{eq:rMS2} can be decomposed into $G$ parallel MIMO AWGN channels as
  \begin{align}
  \mathbf r_{\bar{\mathcal M}_g} = \bar{\mathbf H}_g \mathbf x_{\bar{\mathcal Q}_g} + \mathbf z_{\bar{\mathcal M}_g}, \ g=1,\cdots, G, \label{eq:rMgSufficientAoA}
  \end{align}
  where $\mathbf r_{\bar{\mathcal M}_g}, \mathbf z_{\bar{\mathcal M}_g}\in \mathbb{C}^{|\bar{\mathcal M}_g|\times 1}$ and $\mathbf x_{\bar{\mathcal Q}_g}\in \mathbb{C}^{|\bar{\mathcal Q}_g|\times 1}$  denote the sub-vectors of $\mathbf r_{\mathcal M_S}$,  $\mathbf z_{\mathcal M_S}$ and $\mathbf x_{\mathcal Q_S}$ in \eqref{eq:rMS2}, respectively; and $\bar{\mathbf H}_g\triangleq \sum_{l\in \mathcal L_g} \alpha_l \mathbf a_{R,\bar{\mathcal M}_g}(\phi_{R,l})\mathbf a_{T,\bar{\mathcal Q}_g}^H (\phi_{T,l})$ denotes the corresponding MIMO channel matrix for group $g$. The channel capacity of \eqref{eq:rMgSufficientAoA} is then achieved by the  eigenmode transmission over each of the $G$ parallel MIMO channels, which have smaller dimension and hence require  lower  complexity as compared to the original channel in \eqref{eq:rMS2} without path grouping.

\begin{figure}
\centering
\includegraphics[scale=1]{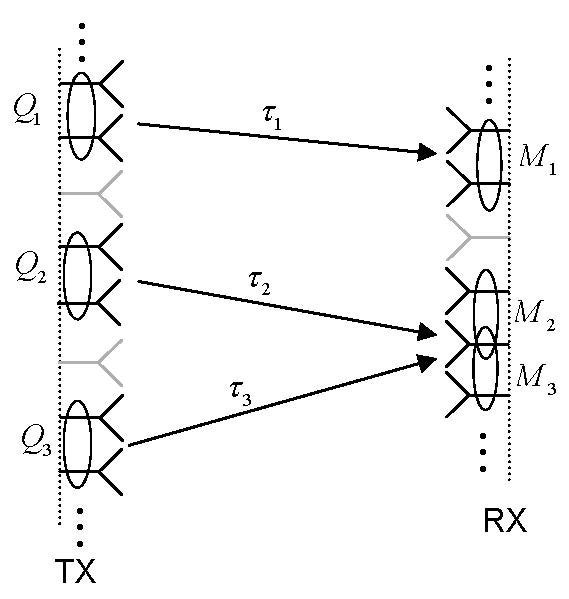}
\caption{An illustration of the effective channel in mmWave lens MIMO system with sufficiently separated AoDs. The path delays are labeled for each link.  Gray antennas represent those with negligible power and hence can be deactivated. Note that path $2$ and path $3$  are grouped since they have similar AoAs at Rx (but different AoDs at Tx).}\label{F:ModelWellSeparatedAoDs}
\end{figure}

\subsubsection{Sufficiently Separated AoDs}\label{sec:sufficientAoDs}
Next, we consider the case with sufficiently  separated AoDs for all paths such that $|\tphi_{T,l}-\tphi_{T,l'}|>2\Delta/\tD_T$, or $\mathcal Q_l \cap \mathcal Q_{l'}=\emptyset$, $\forall l\neq l'$, but  with  possibly close AoAs for certain paths. This may correspond to the downlink transmission with  accurate AoD resolution ($\tD_T\gg 1$) at the base station transmitter, but with only moderate AoA resolution at the mobile terminal receiver. In this case, $\{\mathcal Q_l\}_{l=1}^L$ form a disjoint partition for the transmitting antenna subset $\mathcal Q_S$, and the input-output relationship in \eqref{eq:rMS} can be re-written as
\begin{align}
\mathbf r_{\mathcal M_S}(t) = \sum_{l=1}^L \alpha_l \mathbf a_{R,\mathcal M_S}(\phi_{R,l}) \mathbf a_{T,\mathcal Q_l}^H (\phi_{T,l}) \mathbf x_{\mathcal Q_l}(t-\tau_l)+\mathbf z_{\mathcal M_S}(t).\label{eq:rMS3}
\end{align}
The expression in \eqref{eq:rMS3} shows that the signals sent by each transmitting antenna arrive at the receiver only via  one of the  multi-paths, as illustrated in Fig.~\ref{F:ModelWellSeparatedAoDs}. This thus provides the opportunity for path delay pre-compensation at the transmitter by setting the transmitted signal as $\mathbf x_{\mathcal Q_l}(t)=\mathbf x_{\mathcal Q_l}'(t+\tau_l)$, $\forall l$. As a result, \eqref{eq:rMS3} can be equivalently written as \eqref{eq:rMSSufficientAoDs} shown at the top of the next page.
\begin{figure*}
\begin{align}
\mathbf r_{\mathcal M_S}=\underbrace{\left[\begin{matrix}  \alpha_1 \mathbf a_{R,\mathcal M_S}(\phi_{R,1})\mathbf a_{T,\mathcal Q_1}^H(\phi_{T,1})  & \cdots  &  \alpha_L \mathbf a_{R,\mathcal M_S}(\phi_{R,L})\mathbf a_{T,\mathcal Q_L}^H(\phi_{T,L})
\end{matrix} \right]}_{\mathbf H_S} \underbrace{\left[\begin{matrix}\mathbf x_{\mathcal Q_1}' \\ \vdots \\ \mathbf x_{\mathcal Q_L}' \end{matrix}\right]}_{\mathbf x_{\mathcal Q_S}'}+ \mathbf z_{\mathcal M_S}. \label{eq:rMSSufficientAoDs}
\end{align}
\hrule
\end{figure*}

Similar to the previous subsection, \eqref{eq:rMSSufficientAoDs} shows that with sufficiently separated AoDs, the lens MIMO system is  equivalent to a $|\mathcal M_S|\times |\mathcal Q_S|$  MIMO AWGN channel. This holds regardless of narrow-band or wide-band communications.  The channel capacity of \eqref{eq:rMSSufficientAoDs} is achievable by eigenmode transmission with WF power allocation based on the equivalent channel matrix $\mathbf H_S$. Similar to Section~\ref{sec:sufficientAoAs}, by exploiting the channel sparsity of  $\mathbf H_S$, we can design a low-complexity capacity-achieving scheme by employing the {\it AoA-based path-grouping} at the receiver side. Specifically, the $L$ signal paths are partitioned into $G\leq L$ groups such that paths $l$ and $l'$ belong to the same group if their supporting receiving antenna subsets have non-empty overlapping, i.e., $\mathcal M_l \cap \mathcal M_{l'}\neq \emptyset$. Denote by $\mathcal L_g\subseteq \{1,\cdots, L\}$, $g=1,\cdots, G$, the subset containing all paths in group $g$. For instance, for the system shown in Fig.~\ref{F:ModelWellSeparatedAoDs}, we have $G=2$ and $\mathcal L_1=\{1\}$ and $\mathcal L_2=\{2,3\}$.    In addition, denote by $\mathcal{\bar Q}_{g} \triangleq \bigcup_{l\in \mathcal L_g} \mathcal Q_l$ and $\bar {\mathcal M}_g\triangleq \bigcup_{l\in \mathcal L_g} \mathcal M_l$, $g=1,\cdots, G$, the supporting transmitting and receiving antenna subsets for all paths in group $g$, respectively. Similar to   Section~\ref{sec:sufficientAoAs}, the input-output relationship in \eqref{eq:rMSSufficientAoDs} can then be decomposed into $G$ parallel MIMO AWGN channels as
  \begin{align}
  \mathbf r_{\bar{\mathcal M}_g} = \bar{\mathbf H}_g \mathbf x_{\bar{\mathcal Q}_g}' + \mathbf z_{\bar{\mathcal M}_g}, \ g=1,\cdots, G, \label{eq:rMg}
  \end{align}
  where $\mathbf r_{\bar{\mathcal M}_g}, \mathbf z_{\bar{\mathcal M}_g}\in \mathbb{C}^{|\bar{\mathcal M}_g|\times 1}$ and $\mathbf x_{\bar{\mathcal Q}_g}'\in \mathbb{C}^{|\bar{\mathcal Q}_g|\times 1}$  denote the sub-vectors of $\mathbf r_{\mathcal M_S}$,  $\mathbf z_{\mathcal M_S}$ and $\mathbf x_{\mathcal Q_S}'$ in \eqref{eq:rMSSufficientAoDs}, respectively; and $\bar{\mathbf H}_g\triangleq \sum_{l\in \mathcal L_g} \alpha_l \mathbf a_{R,\bar{\mathcal M}_g}(\phi_{R,l})\mathbf a_{T,\bar{\mathcal Q}_g}^H (\phi_{T,l})$ denotes the corresponding MIMO channel matrix for group $g$. The channel capacity of \eqref{eq:rMg} is then achieved by the eigenmode transmission over each of the $G$ parallel MIMO channels each with   reduced size. %, which have smaller dimension and hence lower  complexity than the original channel in \eqref{eq:rMSSufficientAoDs} without path grouping.
  %In the favorable scenario where the AoAs are also sufficiently separated, i.e., $\mathcal M_l \cap \mathcal M_{l'}=\emptyset$, $\forall l'\neq l$, it can be shown that the above path-grouping based scheme reduces to the simple PDM scheme proposed in Section~\ref{sec:PDM}.

\subsection{Numerical Results}
In this subsection, we evaluate the performance of the proposed PDM  in a wide-band mmWave lens MIMO system  by numerical examples. We assume that the lens apertures at the transmitter and receiver are $A_T=100$ and $A_R=50$, respectively,  and the azimuth lens dimensions are $\tD_T=20$ and $\tD_R=10$, respectively. Accordingly, the number of transmitting and receiving antennas in the lens MIMO systems are $Q=41$ and $M=21$, respectively. For the benchmark MIMO system based on the conventional UPAs, the number of transmitting and receiving antennas are set as $Q_\upa=400$ and $M_\upa=200$, respectively,  for achieving the same antenna apertures as the lens MIMO system.  For both the lens MIMO and   UPA-based MIMO systems, antenna selections are applied by  assuming that the number of RF chains at the transmitter and receiver are $\Mrf=\Qrf=2\Delta L$, where $L$ is the number of multi-paths and $\Delta$ is a design parameter to achieve a reasonable balance between performance and RF chain cost. We set $\Delta=1$ in this example. For the lens MIMO system, the AoA/AoD based antenna selection method  given in \eqref{eq:Ml} and \eqref{eq:Ql} are applied at the receiver and transmitter, respectively.  However, since the optimal antenna  scheme for the UPA-based MIMO-OFDM system is unknown in general, we adopt the power-based antenna selection due to its simplicity and good performance \cite{370}.  We assume that the mmWave channel has $L=3$ paths with AoDs given by $\phi_{T,l}\in \{-15^\circ, 10^\circ, 45^\circ\}$, which are sufficiently separated based on the criterion specified in Section~\ref{sec:sufficientAoDs}.
On the other hand, the AoAs of the $L$ paths are assumed to be equally spaced in the interval $[-\Phi_R/2, \Phi_R/2]$, with $\Phi_R$   referred to as the AoA angular spread. Furthermore, the %complex-valued path gains $\{\alpha_l\}_{l=1}^L$ are modeled exactly in the same way as in Section~\ref{sec:capacityComp}.  The
 maximum   multi-path delay is assumed to be $T_m=100$ ns and the total available bandwidth is $W=500$ MHz, which is divided into $N=512$ sub-carriers  for the UPA-based MIMO-OFDM. The CP length for the OFDM is set as $100$ ns.

\begin{figure}
\centering
\includegraphics[scale=0.6]{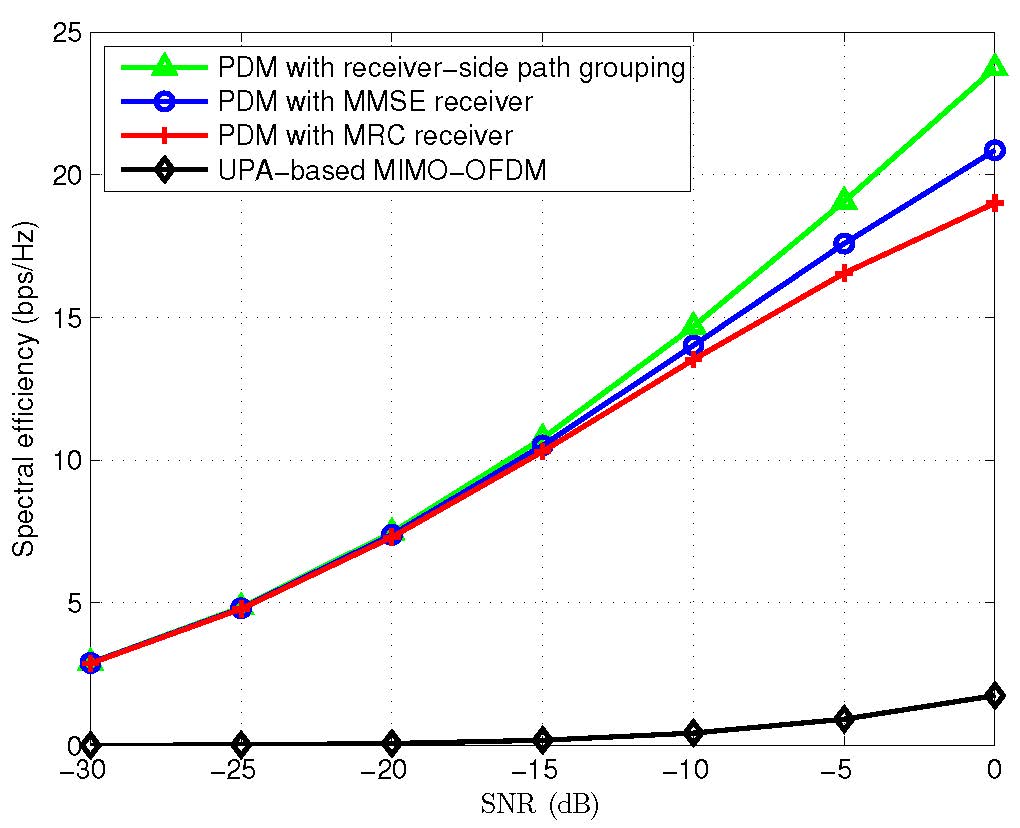}
\caption{Average spectrum efficiency achieved by various schemes  in wide-band mmWave MIMO communication with antenna selection. The number of transmitting/receiving RF chains are $\Mrf=\Qrf=6$. AoA angular spread is $\Phi_R=150^\circ$.}\label{F:RateVsSNRWBAS150}
\end{figure}

\begin{figure}
\centering
\includegraphics[scale=0.6]{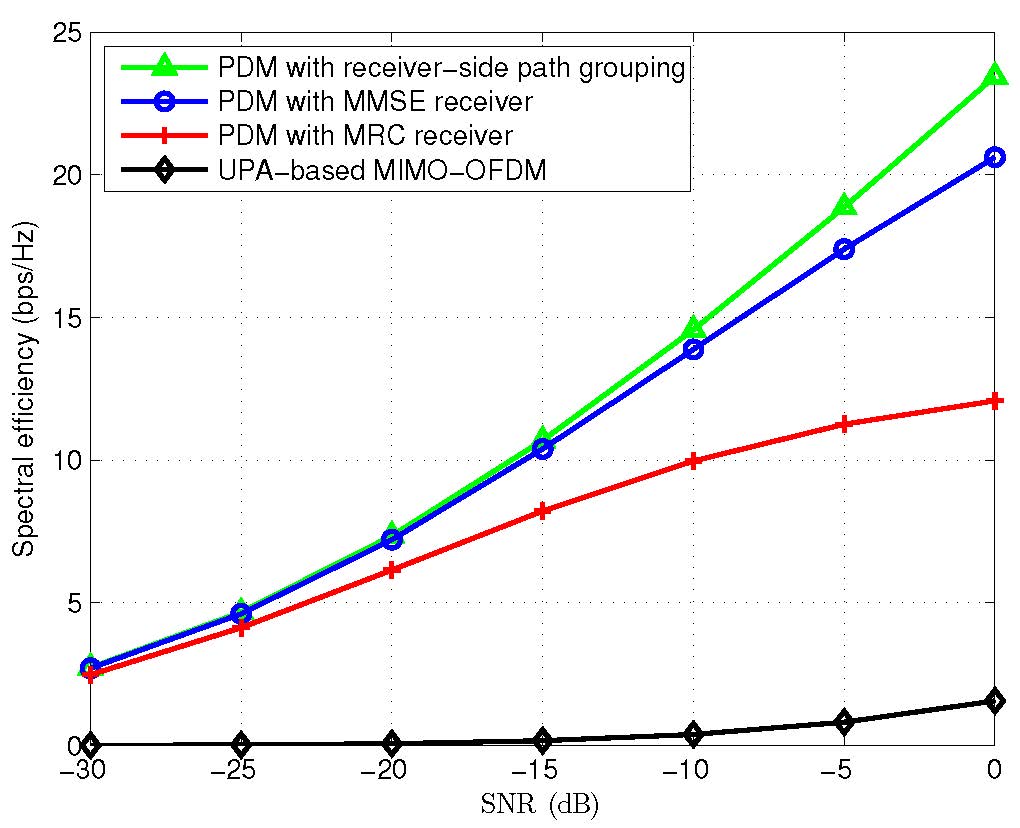}
\caption{Average spectrum efficiency achieved by various schemes in wide-band mmWave MIMO communication with antenna selection. The number of transmitting/receiving RF chains are $\Mrf=\Qrf= 6$. AoA angular spread is $\Phi_R=10^\circ$.}\label{F:RateVsSNRWBAS10}
\end{figure}

In Fig.~\ref{F:RateVsSNRWBAS150},  the spectrum efficiency achieved by different schemes is shown for the mmWave communication with AoA angular spread $\Phi_R=150^\circ$. Note that for simplicity the power allocation $\{p_l\}_{l=1}^L$ for the PDM   with MRC and MMSE receive  beamforming is obtained via WF by assuming $L$ parallel SISO channels with power gains $\{|\alpha_l|^2A_RA_T\}_{l=1}^L$. It is observed from Fig.~\ref{F:RateVsSNRWBAS150} that the  UPA-based MIMO-OFDM gives rather poor performance, which is expected due to the limited array gain with the small number of antennas selected. In contrast, the lens MIMO systems with  the three proposed PDM schemes achieve significant rate improvement over the UPA-based MIMO-OFDM with the same number of RF chains used or antennas selected. Moreover, Fig.~\ref{F:RateVsSNRWBAS150} shows that in the low-SNR regime,  PDM with the simple MMSE and MRC receive beamforming achieves the same performance as that   with path grouping, which is expected due to the negligible inter-path interference in the low-SNR regime. While as the SNR increases, the  three  PDM schemes show more different performances  due to their different interference mitigation capabilities. The performance gaps are more pronounced for systems with smaller AoA separations, as shown in Fig.~\ref{F:RateVsSNRWBAS10} for $\Phi_{R}=10^\circ$ as compared to Fig.~\ref{F:RateVsSNRWBAS150} for $\Phi_{R}=150^\circ$. This implies the necessity of more sophisticated interference mitigation techniques (such as path grouping) for   PDM   when the paths are  severely coupled with each other due to the limited AoA/AoD separations.

\section{Conclusion and Future Work}\label{sec:Conclu}
In this paper, we studied the use of lens antenna arrays for mmWave MIMO communications. The   array response of the lens antenna array  was derived and  compared with that of conventional UPA without the lens. We showed that the proposed  lens antenna array   significantly reduces the signal processing complexity and  RF chain cost as compared to the conventional UPA in mmWave MIMO communications, and yet without notable performance degradation.  We proposed a new low-complexity  MIMO spatial multiplexing technique called PDM,  for both narrow-band and wide-band communications. Analytical results showed that the PDM scheme is able to achieve perfect ISI rejection as long as the AoAs or AoDs are sufficiently separated, thanks to the energy focusing capability of the lens antenna. Finally, for cases with insufficient  AoA/AoD separations, a simple path grouping technique was proposed for PDM to mitigate inter-path interference more effectively.

There are a number of interesting directions that are  worthy of future investigation, which are briefly discussed as follows.
\begin{itemize}
\item {\it Elevation AoAs/AoDs:} For systems with non-negligible  elevation AoAs/AoDs, the  array configuration of the lens antenna arrays needs to be refined. In this case, the antenna elements should be generally placed on the {\it focal surface} of the EM lens to exploit the elevation angular dimension as well, instead of on the focal arc only as considered in this paper. As a result, the signal multi-paths  can be further differentiated with the additional elevation AoA/AoD dimension.
\item {\it Multi-User Systems:} The PDM for the point-to-point mmWave MIMO communication can be extended to the general {\it path division multiple access} (PDMA) for multi-user mmWave systems, by which a number of users with well separated AoAs/AoDs can be simultaneously served with   low-complexity and low-cost transceiver designs. The transmission scheduling of users based on their  AoAs/AoDs   is also worth investigating.
\item {\it Channel Estimation:} In this paper, we assume perfect channel state information at both the transmitter and receiver, while in practical mmWave systems such knowledge needs to be efficiently obtained via well-designed channel training/estimation/feedback schemes.  For mmWave MIMO communications with conventional arrays, channel estimation is a challenging task due to the large-antenna dimension as well as the low SNR before beamforming is applied \cite{574,573,579}; whereas with lens antenna arrays, by exploiting its energy focusing as well as  the multi-path sparsity of mmWave channels, the effective channel dimension is significantly reduced and the pre-beamforming SNR is greatly enhanced. Therefore, channel knowledge can be obtained far more efficiently as compared to conventional arrays, which deserves    further study.
\end{itemize}

 \appendices

 \section{Proof of Lemma~\ref{lemma:response}}\label{A:lensArray}
 To derive the array response of the proposed lens antenna array   given in Lemma~\ref{lemma:response}, we  first present the fundamental principle of operation for EM lenses.  EM lenses are fundamentally  similar to optical lenses, which are able to   alter the propagation directions of the EM rays to achieve energy focusing or beam collimation.

\begin{figure}
\centering
\includegraphics[scale=0.6]{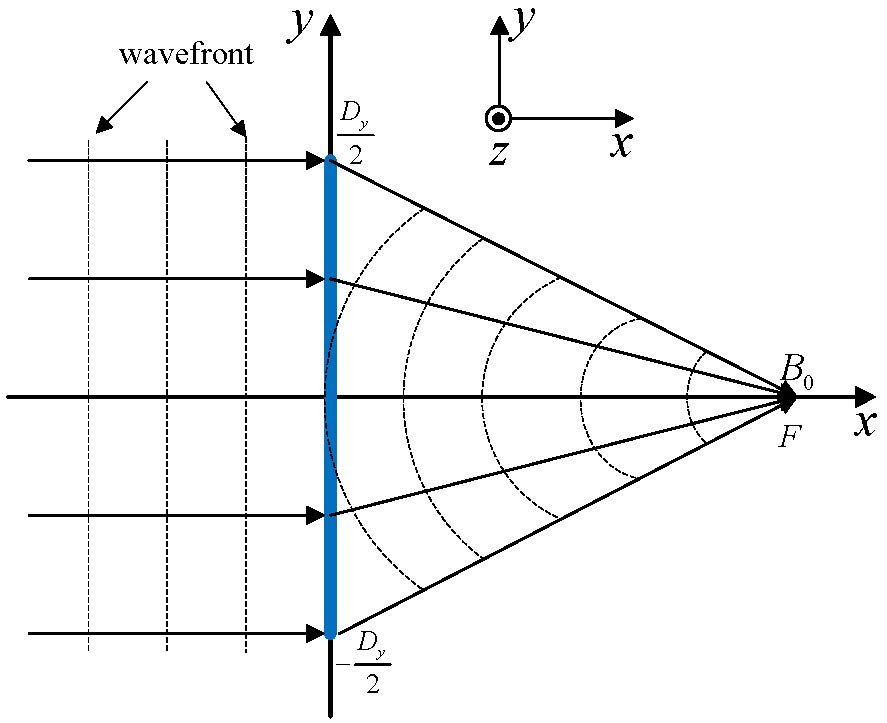}
\caption{Top view of a planar EM lens placed in the y-z plane with focal point $B_0(F,0,0)$ for normal incident plane waves.}\label{F:EMlens}
\end{figure}

 Fig.~\ref{F:EMlens} shows a planar EM lens of size $D_y\times D_z$ placed in the y-z plane and centered at the origin. Denote by $B_0$ with coordinate $(F, 0,0)$ the focal point of the lens for normal incident plane waves, where $F$ is known as the focal length.  The main mechanism to achieve energy focusing at $B_0$ is to design the phase shift profile $\Phi(y,z)$, which represents the phase delay provided by the spatial phase shifters (SPS) of the lens at any point $(0,y,z)$ on the lens's aperture, such that all rays with normal incidence arrive at $B_0$ with  identical phase for constructive superposition \cite{554}. We thus have
\begin{align}
\Phi(y,z)+ & k_0 d(y,z,B_0)=\Phi_0,\notag \\
& \forall (y,z)\in \left[-\frac{D_y}{2}, \frac{D_y}{2} \right] \times \left[-\frac{D_z}{2}, \frac{D_z}{2} \right],
\end{align}
where $k_0=2\pi/\lambda$ is the free-space wave number of the incident wave, with $\lambda$ denoting the free-space wavelength, $d(y,z,B_0)=\sqrt{F^2 + y^2 + z^2}$ is the distance between the point $(0,y,z)$ on the lens's aperture and the focal point $B_0$, and $\Phi_0$ is a positive constant denoting the common phase delay from the lens's input aperture to the focal point $B_0$. The phase shift profile is then designed to be
\begin{align}
\Phi(y,z)= \Phi_0 & -  k_0\sqrt{F^2 + y^2 + z^2}, \notag \\
 & \forall (y,z)\in \left[-\frac{D_y}{2}, \frac{D_y}{2}\right] \times \left[-\frac{D_z}{2}, \frac{D_z}{2}\right]. \label{eq:Phi}
\end{align}

As can be seen from \eqref{eq:Phi}, due to the different propagation distances from the lens's aperture to $B_0$, the phase shift profile varies across the lens apertures with different $y$ and $z$ values. In general, larger phase delay needs to be provided by the SPS located in the center of the lens than those on the edge.

\begin{figure}
\centering
\includegraphics[scale=0.6]{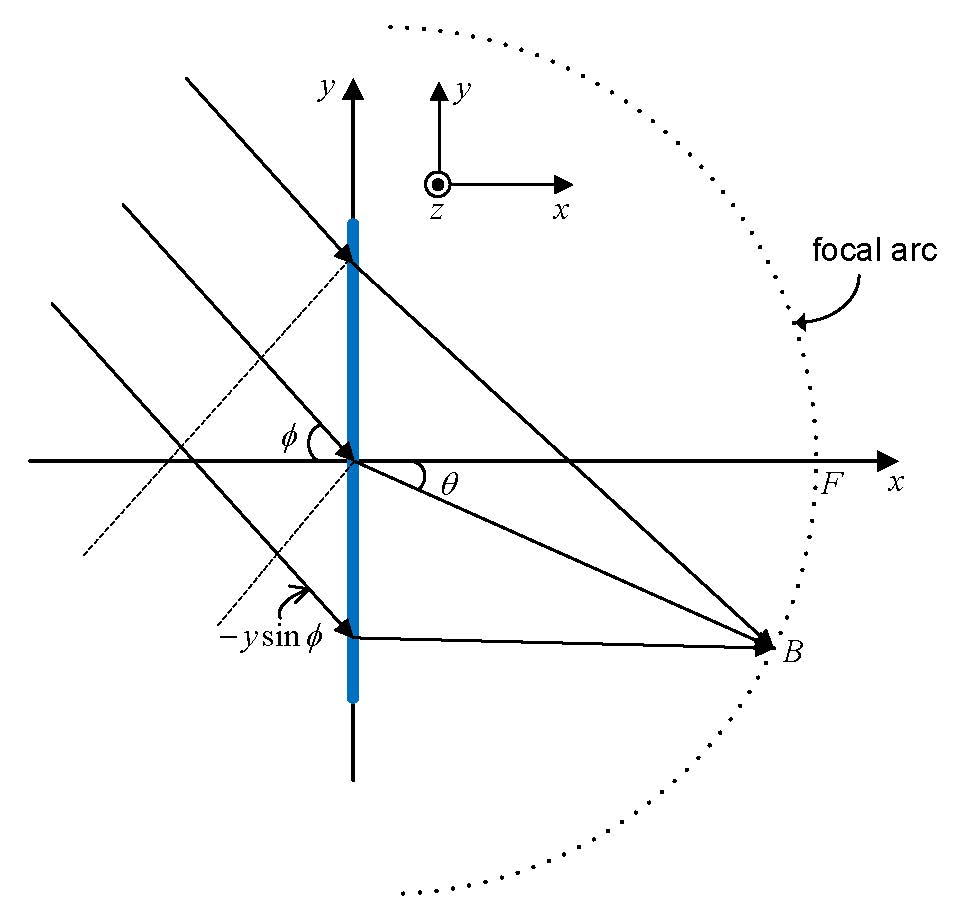}
\caption{A planar EM lens placed in the y-z plane with oblique incident plane wave of azimuth AoA $\phi$.}\label{F:EMlensOblique}
\end{figure}

With the phase shift profile designed in \eqref{eq:Phi} to achieve focal point $B_0$ for normal incident wave, the resulting phase delay from the lens's input aperture $(0,y,z)$ to an arbitrary point $B(x_B,y_B,z_B)$ is then given by
\begin{align}
\psi(y,z,B)=\Phi(y,z)+k_0d(y,z,B), \label{eq:psi}
\end{align}
where $d(y,z,B)=\sqrt{x_B^2+(y_B-y)^2+(z_B-z)^2}$ denotes the distance from the point $(0,y,z)$ on the lens to point $B$. Of particular interest is the field distribution on the {\it focal arc} of the lens, which is defined as the arc on the x-y plane with a distance $F$ from the lens center, as shown in Fig.~\ref{F:EMlensOblique}. Let $B(F \cos\theta, -F \sin \theta,0)$ be a point on the focal arc parameterized by angle $\theta\in [-\frac{\pi}{2}, \frac{\pi}{2}]$. With \eqref{eq:Phi} and \eqref{eq:psi}, we have
{\small
\begin{align}
\hspace{-3ex} \psi(y,z,\theta)&=\Phi_0-k_0\sqrt{F^2 + y^2 + z^2} + k_0 \sqrt{F^2 + y^2 + z^2 +2yF\sin\theta} \label{eq:exact}\\
&\approx \Phi_0 +k_0 y  \sin \theta,\label{eq:approx}
\end{align}
}
where \eqref{eq:approx} follows from the first-order Taylor approximation and the assumption that $F\gg D_y, D_z$.

%We first consider the case when the lens is applied at the receiver for receiving signals from different directions.
%Without loss of generality, we derive the array response by assuming that the lens is applied at the receiver for receiving signals from different directions.
 Let $s(y,z)$ denote the incident signal arriving at the lens's input aperture. Due to the linear superposition principle, the resulting signal on the focal arc of the lens can then be expressed as
\begin{align}
r(\theta)& = \int_{-D_z/2}^{D_z/2}\int_{-D_y/2}^{D_y/2} s(y,z) e^{-j\psi(y,z,\theta)} dydz  \\
&= e^{-j\Phi_0}D_z \int_{-D_y/2}^{D_y/2} s(y)e^{-j\frac{2\pi}{\lambda} y \sin\theta}dy, \ \theta \in \left[ -\frac{\pi}{2}, \frac{\pi}{2}\right], \label{eq:rtheta}
\end{align}
where in \eqref{eq:rtheta}, we have assumed that $s(y,z)=s(y)$, $\forall (y,z)\in \left[-\frac{D_y}{2}, \frac{D_y}{2} \right] \times \left[-\frac{D_z}{2}, \frac{D_z}{2} \right]$, which is true for uniform incident plane waves with negligible elevation AoAs. For notational convenience, we assume that $\Phi_0=2n\pi$ for some integer $n$, so that it can be ignored in \eqref{eq:rtheta}. Furthermore, by defining
\begin{align}
\tD=\frac{D_y}{\lambda}, \quad \ty=\frac{y}{\lambda}, \quad \ttheta=\sin(\theta),
\end{align}
 the relationship in \eqref{eq:rtheta} can be equivalently written as
\begin{align}
r(\ttheta)= D_z \int_{-\tD/2}^{\tD/2} \ts(\ty)e^{-j2\pi \ttheta \ty }d\ty, \quad \ttheta\in [-1, 1], \label{eq:fourier}
\end{align}
where $\ts(\ty)$ with $\ty\in \left[-\tD/2, \tD/2 \right]$ is a linear scaling of the arriving signal $s(y)$ given by $\ts(\ty)\triangleq \lambda s(\lambda \ty)$.

It is interesting to observe from \eqref{eq:fourier} that with the spatial phase shifting provided by the EM lens, the resulting signal at the focal arc of the lens can be represented  as the {\it Fourier transform} of the arriving signal $\ts(\ty)$ at the lens's input aperture, with $\ttheta\in [-1,1]$ and $\ty\in \left[-\tD/2, \tD/2 \right]$ in \eqref{eq:fourier} referred to as the {\it spatial frequency} and the {\it spatial time}, respectively.

 For  uniform incident plane waves with azimuth AoA $\phi$, or equivalently with spatial frequency $\tphi=\sin(\phi)$, as shown in Fig.~\ref{F:EMlensOblique}, we have
$s(y)=\frac{1}{\lambda \sqrt{D_yD_z}} x_0(\phi)e^{j\frac{2\pi}{\lambda}y\sin(\phi)}$,
or equivalently
\begin{align}
\ts(\ty)=\frac{1}{\sqrt{D_yD_z}} x_0(\phi)e^{j2\pi\ty\tphi}, \label{eq:tsty}
\end{align}
where $x_0(\phi)$ is the input  signal arriving at the lens center with AoA $\phi$, and $\sqrt{D_yD_z}$ is a normalization factor to ensure that the total power captured by the lens is proportional to its effective aperture $A\triangleq D_y D_z/\lambda^2$. By substituting \eqref{eq:tsty} into \eqref{eq:fourier}, we have
\begin{align}
r(\ttheta)= x_0(\phi) \sqrt{A} \sinc\left(\tD(\ttheta-\tphi) \right), \quad \ttheta\in [-1, 1]. \label{eq:rttheta}
\end{align}
 It then follows from \eqref{eq:rttheta} that the effective lens response on its focal arc for incident plane waves with AoA $\phi$ (or spatial frequency $\tphi$) is
\begin{align}
a_{\ttheta}(\phi) = \sqrt{A} \sinc(\tD(\ttheta-\tphi)),\ \ttheta \in [-1, 1]. \label{eq:response}
\end{align}

%Several observations can be made from \eqref{eq:rttheta} and \eqref{eq:response}. First, for an incident plane wave with AoA $\phi$, a signal peak with power intensity proportional to the lens aperture $A$ is achieved at one particular point on the focal arc with $\theta=\phi$, or $\ttheta=\tphi$. Second, for $\ttheta-\tphi=n/ \tDy$ for some non-zero integer $n$, or $\ttheta-\tphi\gg 1/ \tDy$, the lens response $a(\ttheta,\tphi)$ is almost negligible. Thus, the quantity $\Delta \tphi\triangleq 1/\tDy$ is termed as the {\it spatial frequency resolution} of the lens, which corresponds to the angle resolution $\Delta \phi=\sin^{-1}(1/\tDy)= 1/\tDy$ for large $\tDy$ \cite{553}.

For the lens antenna array with the $m$th element located at position $B_m(F\cos(\theta_m), -F\sin(\theta_m), 0)$, it follows from \eqref{eq:response} that the array response can be expressed as
\begin{align}
a_{m}(\phi)=\sqrt{A} \sinc \left(\tD(\sin \theta_m-\sin\phi)\right), \ \forall m. \label{eq:aRm}
\end{align}

%With the so-called {\it critical antenna spacing}, the antenna elements are positioned on the focal arc so that the spatial frequency $[-1, 1]$ is uniformly sampled by
In particular, with the critical antenna spacing specified in \eqref{eq:sinThetam}, the array response in \eqref{eq:aRm} reduces to
 \begin{align}
 a_{m}(\phi)=\sqrt{A}\sinc\left(m-\tD \sin \phi \right), \ \forall m. \label{eq:aRmcritical}
 \end{align}

This completes the proof of Lemma~\ref{lemma:response}.

\bibliographystyle{IEEEtran}
\bibliography{IEEEabrv,IEEEfull}
\end{document}